\documentclass[12pt]{article}

\usepackage{graphicx}
\usepackage{times}
\usepackage[section]{placeins}
\usepackage{longtable}
\usepackage{url}
\usepackage{tabularx}

\usepackage{filecontents}

\usepackage{tikz-cd}

\newcommand{\maybe}[1]{}

\newcolumntype{L}[1]{>{\raggedright\arraybackslash}p{#1}}

\usepackage{xr} 

\title{Inference under Superspreading: 
Determinants of SARS-CoV-2 Transmission in Germany} 

\author
{Patrick Schmidt
\\
\normalsize{
	University of Zurich}\\
\normalsize{
E-mail:  PatrickWolfgang.Schmidt@uzh.ch}
}

\date{}

\begin{document} 

\maketitle 

\begin{abstract} 
Superspreading complicates the study of SARS-CoV-2 transmission. I propose a model for aggregated case data that accounts for superspreading and improves statistical inference. In a Bayesian framework, the model is estimated on German data featuring over 60,000 cases with date of symptom onset and age group. 
Several factors were associated with a strong reduction in transmission: public awareness rising, testing and tracing, information on local incidence, and high temperature. Immunity after infection, school and restaurant closures, stay-at-home orders, and mandatory face covering were associated with a smaller reduction in transmission. The data suggests that  public distancing rules increased transmission in young adults. Information on local incidence was associated with a reduction in transmission of up to 44\% (95\%-CI: [40\%, 48\%]), which suggests a prominent role of behavioral adaptations to local risk of infection. Testing and tracing reduced transmission by 15\% (95\%-CI: [9\%,20\%]), where the effect was strongest among the elderly. Extrapolating weather effects, I estimate that transmission increases by 53\% (95\%-CI: [43\%, 64\%]) in colder seasons.
\end{abstract}

\newpage
\subsection*{Introduction}
At the point of writing this article, the world records a million deaths associated with Covid-19 and over 30 million people have tested positive for the SARS-CoV-2 virus. Societies around the world have responded with unprecedented policy interventions and changes in behavior. The reduction of transmission has arisen as a dominant strategy to prevent direct harm from the newly emerged virus. Yet, quantitative evidence on the determinants of transmission remains scarce. 

Individual variation in transmission (dispersion or potential for so called superspreading \cite{lloyd2005superspreading}) is one of the reasons why the collection of adequate evidence is challenging. Overdispersion (high likelihood of superspreading) is well documented for SARS-CoV-1 \cite{riley2003transmission,lipsitch2003transmission} and is shown to be a dominant feature of SARS-CoV-2 (see also \cite{adam2020clustering}). Under superspreading single cases convey little information as transmission kinetics are driven by few outliers. Thus, conclusions drawn from anecdotal evidence with few primary cases (e.g., \cite{hendrix2020absence} and \cite{khan2020transmission}) are subject to substantial uncertainty.

In the absence of sufficiently large and detailed contact tracing data, some research has relied on surveillance data with aggregated cases \cite{sehra2020maximum,salje2020estimating,CHERNOZHUKOV2020} or deaths \cite{flaxman2020estimating,CHERNOZHUKOV2020}. However, such data provide additional challenges to inference, including a time lag in reporting, under-reporting of cases, and a lack of established methods to account for superspreading. 

I address those issues with the following three points. First, by aggregating cases based on \emph{symptom onset} instead of \emph{reporting date}, the timing of transmission is estimated more sharply. Second, I account for underreporting by modelling infections as compartmentalized \emph{unobservable} latent process. Age compartments allow to identify age specific growth rates if reporting differs between age groups. Finally, the paper proposes a \emph{probabilistic} infection process, that extends well-established models for superspreading \cite{lloyd2005superspreading} to aggregated case counts. \maybe{Under elementary assumptions, the distribution of aggregated case counts is derived. This allows probabilistic statements on reproductive numbers and their determinants based on aggregated case counts while accounting for the fact that compartments with low(high) case counts provide more(less) noisy signals in a principled manner. }

\maybe{In summary, my approach allows to compartmentalize data, which improves identification of growth rates, while aggregating all information for reliable inference under superspreading. The study illustrates the importance of standardized public surveillance data that contains information on age, location, and symptom onset.
}

I apply the model to German surveillance data and find that transmission was reduced predominantly by public awareness rising and voluntary behavioral adaptations to publicly reported local incidence. Furthermore, extrapolation suggests strong seasonal effects and potential for testing and tracing.

\maybe{The inference on weather effects, testing delay, contact tracing and information can be argued to be more reliable, as other factors were less volatile after the first wave during a time of reopening. Transmission is negatively correlated with average daily temperature and positively with relative humidity, suggesting a seasonal effect of $xx$ to $xx\%$.}

In the following, I introduce the model. Afterwards, the data is described, and estimation results are illustrated and discussed. A supplementary document contains additional detail in Sections \ref{sec:supp_model} through
\ref{supp:additional}.

\subsubsection*{Model}

The model distinguishes between cases and infections. Infections at time $t$ are denoted by $i_t$. Infections are unobserved. Instead, we observe the number of reported cases $c_t$ with symptom onset at time $t$. The full model (Supplementary Section \ref{sec:supp_model}) uses daily data and accounts for timing as standard in the literature \cite{cori2013new,flaxman2020estimating}. In this section, I present a simplified version to illustrate the implications of superspreading for statistical inference. In particular, the generation time (from primary infection to secondary infection) and incubation period (from infection to symptom onset) are fixed to one time step. Further, indices for age and location are dropped. 

\begin{figure}[htp]
	\centering
	\begin{tikzcd}[cells={nodes={draw=gray, circle}}] 
	\dots 	&  i_{t-1} \arrow[rr, "R_t" description, bend left] 
	&
	& |[draw=black,thick,fill=lightgray]| i_t 
	\arrow[rrd, "r_{t+1}" description,bend left = 5, shift right=2] 
	\arrow[rr, "R_{t+1}" description, bend left]
	& & i_{t+1}          &   \dots           \\
	&    \dots  &  & c_t  &    & c_{t+1} &  \dots\\
	\end{tikzcd}
	\caption{Model illustration. All connections to a new infection at time $t$ are shown. A primary infection at time $t-1$ infects on average $R_t$ secondary infections. A person infected at time $t$ develops symptoms at time $t+1$ and is reported as case with probability $r_{t+1}$.}
	\label{fig:model_short}
\end{figure}
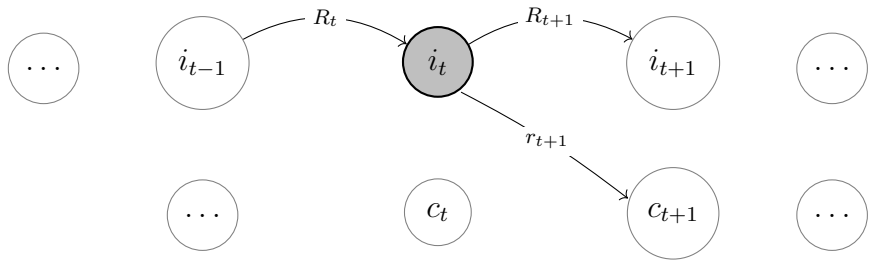

This simplified model is illustrated in Figure \ref{fig:model_short}. A primary infection causes new infections in the following time interval. The average number of secondary infections caused by one primary infection is denoted by the \emph{reproductive number} $R_t$. The probability of an infection being reported as a case is denoted by the \emph{reporting rate} $r_{t}$.

In a popular transmission model \cite{lloyd2005superspreading} secondary infections follow a negative binomial distribution with mean $R_t$ and dispersion $\Psi$, i.e. $NB(R_t,\Psi)$. Dispersion $\Psi$ describes the individual variation in transmission, where small values are associated with high variation. The reproductive number $R_t$ is determined by the conditions that transmissions at time $t$ are subject to. Given those conditions, I assume that the number of secondary infections caused by one primary infection is independent across individuals. This implies for infections $i_t$ that
$$i_t \sim NB(i_{t-1} R_t,i_{t-1} \Psi).$$ 
Thus, infections are more dispersed under a small infection count $i_{t-1}$. For large infection counts $i_{t-1}$ the model converges to a Poisson model with no overdispersion. 

The growth rate $\frac{i_t}{i_{t-1}}$ constitutes a random variable with mean $R_t$ and variance $\frac{R_t (\Psi+R_t)}{i_{t-1}\Psi}$. Variation in the growth rate is caused by variation in the mean $R_t$ and noise influenced by individual dispersion $\Psi$. We can separate the two factors, as the former influences growth rates irrespective of the current infection count, while the latter has less impact for many infections $i_{t-1}$. Empirical weekly growth rates in Germany fit this model and can be used to estimate the dispersion parameter (see Section \ref{supp:variance_growth}).

The model presented in this study deviates from the literature, where overdispersion is often ignored \cite{CHERNOZHUKOV2020,hsiang2020effect,davies2020effects} or assumed constant \cite{flaxman2020estimating}. A constant dispersion of aggregated cases arises under the assumption that the number of secondary infections across primary cases are identical. In this case it follows that $i_t \sim NB(i_{t-1} R_t, \Psi)$. \maybe{The growth rate would have the variance $\frac{R_t}{i_{t-1}} +\frac{R_t^2}{\Psi}$ and the dispersion $\Psi$ affects growth rates independently of the number of currently infectious $i_{t-1}$ and therefore could not be separated from variation of the reproductive number $R_t$.} This alternative assumption of a constant dispersion is inherent, but often unappreciated, in inference based on negative binomial regression, the endemic/epidemic model introduced in \cite{held2005statistical}, and epidemiological models with random effects (e.g., the model in \cite{fisher2020ecological}).

\maybe{In this alternative setting, the variance of newly infected is denoted by $\sigma^2_A= i_{t-1} R_t (1 +\frac{i_{t-1} R_t}{\Psi})$, which is larger than $\sigma^2_i$ if $i_{t-1}>\frac{\Psi}{\Psi}$, and has a variance mean ratio $1+\frac{i_{t-1} R}{\Psi}$ that grows linearly with the aggregated infections $i_{t-1}$.}

\maybe{Under both models the growth rate $\frac{i_t}{i_{t-1}}$ would be a random variable with mean $R_t$. Under the alternative assumption the growth rate has a variance of $\frac{R_t}{i_{t-1}} +\frac{R_t}{\Psi^A}$, whereas the model proposed here arrives at $\frac{R_t}{i_{t-1}} +\frac{R_t}{i_{t-1}\Psi}=\frac{R_t (\Psi+1)}{i_{t-1}\Psi}$.}

The model in this study does not feature infections between compartments. Supporting this simplification, a large scale contact tracing study in India found that most transmissions occur within the same age group \cite{Laxminarayaneabd7672}. I discus extensions with importation in Section \ref{supp:importation}. The remaining parts of the model are standard. Cases $c_t$ constitute a sum of Bernoulli trials that is approximated by a Poisson distribution for computational convenience.

The reproductive number is modelled as a function of covariates, where the effect is assumed to be multiplicative. Each location and age group features a basic reproductive number $R_0$ in the absence of all interventions and under average weather conditions. Effect estimates can be interpreted as the ratio of prevented/added infections at a particular time.

The main obstacles for inference include high correlation between covariates, unobservables, and an unknown reporting rate (see Section \ref{supp:assumptions} for details).

\maybe{The availability of symptom onset data, as opposed to reporting data allows a sharper identification of exact timing effects. While the delay from infection to reporting is highly dependent on current testing and reporting practices, the incubation period is a arguably a constant and well-understood characteristic of the infectious agent \cite{McAloone039652}.
}

\maybe{\cite{goldstein2020effect} argue for the relevance of age-specific transmission dynamics and disparities between incidence counts and prevalence.
}

\subsubsection*{Data}

I apply the model to German surveillance data, which features date of symptom onset, age groups (15-34, 35-59, 60-79, and above 80), and geographical units (county or city). Daily location-specific covariates include 21 policy interventions, average temperature, relative humidity, the ratio of traced infectious, local incidence as known at that point in time, local cumulative incidence, weekday fixed effects, and daily, age and location specific error terms. I classified interventions for 10 German states until May 2020. Estimation is based on cases in 112 counties until mid May 2020, which encompasses more than 60,000 symptomatic cases. See the Supplementary Section \ref{supp:data} for additional details. 

\subsubsection*{Reporting rate}

Reproductive numbers are identified even if only a fraction of infections is ascertained as cases \cite{wallinga2004different}. Importantly, said fraction has to be constant over time. As the likelihood of developing symptoms changes with age (Figure \ref{fig:asymptomatic}), the reporting rate $r_t$ cannot be assumed to be constant across age groups. 

\begin{figure}[htp]
	\centering
	\includegraphics[width=\linewidth]{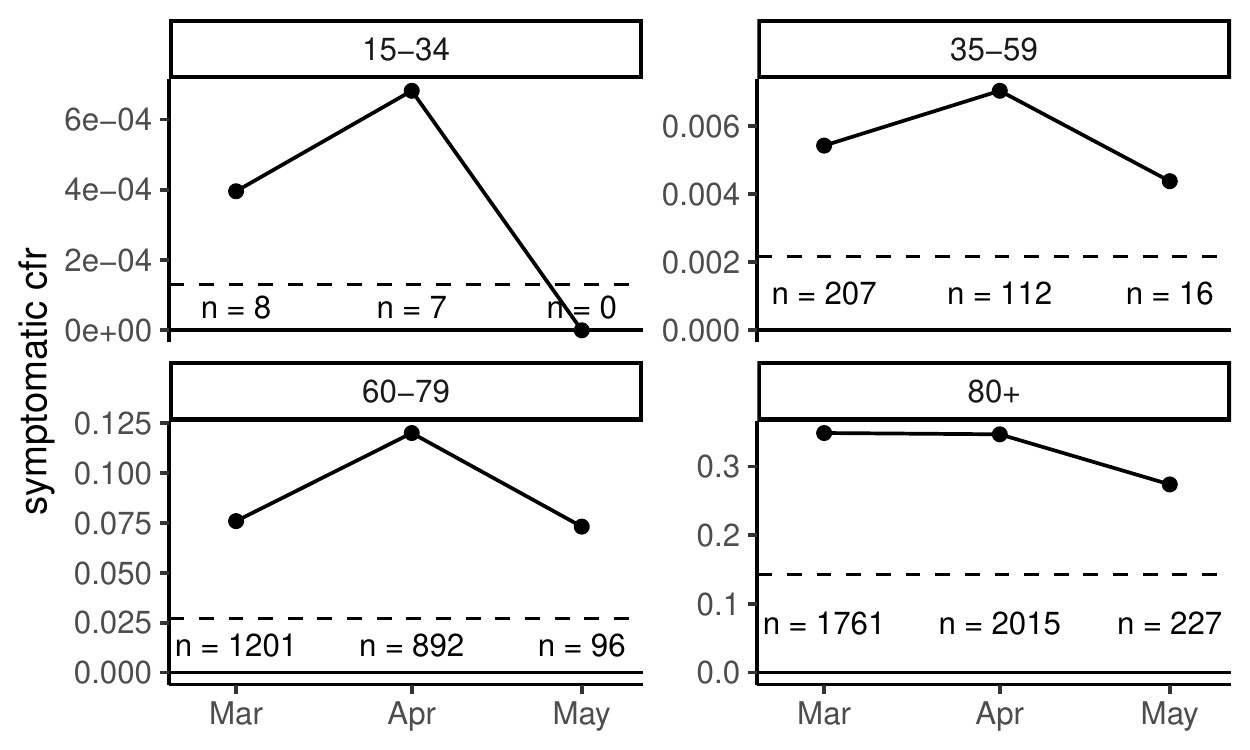}
	\caption{Symptomatic case fatality rate (cfr) over time for different age groups. 
		Ratio of deceased among cases with symptom onset in the model period. Cases without reported symptom onset are not considered. Vertical dashed line denotes age-specific expected infectious fatality rate based on \cite{levin2020assessing}. The number of reported deaths is denoted by $n$ for each age group and month.}
	\label{fig:cfr_main}
\end{figure}

I chose a reporting rate of $25\%$ for the model. \emph{Case fatality rates} can provide some support for this model assumption. If the \emph{infection fatality rate} is constant over time, the case fatality rate identifies changes in the reporting rate. The observed case fatality rate in Figure \ref{fig:cfr_main} does not suggest major changes over time. A comparison to the age-specific infection fatality rates (estimated based on multiple international serological studies \cite{levin2020assessing}) indicates a reporting rate close to $25\%$. Similar reporting rates arise based on first evidence of unpublished serological studies in Germany \cite{rkisero}. Identification of the reproductive number relies on the assumption that symptomatic infections were equally likely to be reported over time within a specific age group and location.

\subsubsection*{Transmission dynamics}

I estimate the model in a Bayesian framework and chose weakly informative priors on transmission characteristics (see Section \ref{sec:supp_priors}). Estimation results are informative about transmission in the context of this study.

The average incubation period was $5.30$ days ($95\%$-CI: $[5.16, 5.45]$). The average generation time was slightly longer with $5.84$ days ($95\%$-CI: $[5.68, 6.01]$). Assuming independence this corresponds to $42\%$ of secondary infections occurring before symptom onset.

The basic reproductive number denotes the average number of secondary infections in the absence of interventions and under average weather conditions. There is no evidence for an age dependent basic reproductive number (Table \ref{tab:transmission}). Population density was positively correlated with the basic reproductive number (Table \ref{tab:meta}, supplementary material).

\begin{table}[htb]
	\centering
	\begin{tabular}{lrlrlrlrl}
		& \multicolumn{2}{c}{\textit{$R_0$}} & \multicolumn{2}{c}{\textit{dispersion $\Psi$}} & \multicolumn{2}{c}{ratio infecting} & \multicolumn{2}{c}{ratio from $20\%$}\\ 
		\cline{2-3} \cline{4-5} \cline{6-7} \cline{8-9}
		age & mean & sd & mean & sd  & mean & sd & mean & sd \\ 
		\hline
		15-34 & 2.61 & 0.24 & 0.75 & 0.16 & 0.47 & 0.04 & 0.69 & 0.03 \\ 
		35-59 & 2.61 & 0.23 & 0.36 & 0.05 & 0.38 & 0.04 & 0.80 & 0.03 \\ 
		60-79 & 2.57 & 0.22 & 5.53 & 2.61 & 0.59 & 0.04 & 0.54 & 0.02 \\ 
		80+ & 2.52 & 0.28 & 3.98 & 2.05 & 0.58 & 0.04 & 0.56 & 0.03 \\ 
		\hline
	\end{tabular}
	\caption{Estimated basic reproductive number $R_0$, dispersion $\Psi$, respective ratio of primary infections actually infecting, and ratio of secondary infections from $20\%$ most infecting primary cases for different age groups. The ratios of secondary infections was computed assuming a constant reproductive number of $1$.}
	\label{tab:transmission}
\end{table}

For the age groups 15-34 and 35-59 we observe high dispersion, with more than half of all cases infecting nobody and $20\%$ of primary cases initiating $70$ to $80\%$ of secondary cases. Older age groups show less tendency for superspreading. Previous work studied dispersion abstracting from age differences. For comparison, I simulated the marginal distribution of secondary infections, which is a mixture negative binomial distribution \cite{furman2007convolution}. Assuming equally distributed infections across age groups, its mean variance ratio is $0.378$ ($95\%$-CI: $[0.33,0.44]$), equivalent to a dispersion parameter of $0.61$ ($95\%$-CI: $[0.49,0.77]$). Contact tracing studies obtained similar results. A study from Hongkong estimated a dispersion parameter of $0.33$ ($95\%$-CI: $[0.14, 0.98]$) or $0.19$ ($95\%$-CI: $[0.13,0.26]$) when assuming a single primary case for an unresolved large cluster. A study from Shenzhen (China) reported $0.58$ ($95\%$-CI: $[0.35, 1.18]$) and a study in two Indian states $0.51$ ($95\%$-CI: $[0.49, 0.52]$).
Global data-sets of outbreaks suggested a higher potential for superspreading with a dispersion parameter of $0.10$ ($95\%$-CI: $[0.05, 0.20]$) \cite{endo2020estimating}. Noteworthy, the aforementioned studies do not account for changing reproductive numbers and are prone to overestimate the prominence of superspreading. Further, clusters might be more likely to be traced, while diffuse community spread is harder to identify.

Other limitations apply to the approach presented here. The model does not allow for overdispersion in reporting and is therefore prone to overestimate the dispersion in transmission. Moreover, changes in the reproductive number might be inadequately modelled. If the reproductive number is over-fitted, individual variation is under-estimated (and vice versa).

\subsubsection*{Effects of policy interventions}

\begin{figure}[htp]
	\centering
	\includegraphics[width=.7\linewidth]{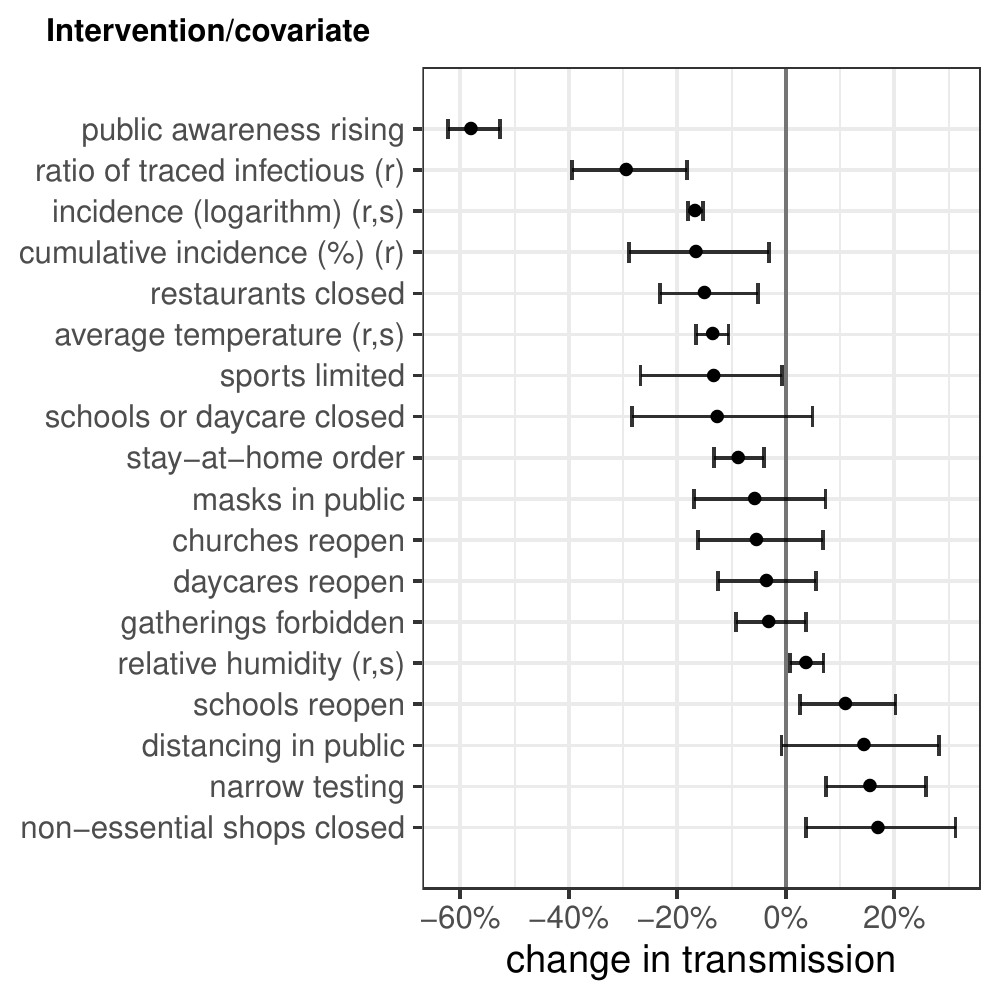}
	\caption{Change in transmission associated with covariates. The plot depicts the mean and $95\%$-CI intervals for age-weighted effect estimates of covariates on the reproductive number $R_t$. Real valued variables (r) and standardized variables (s) are marked.}
	\label{fig:effects_simple}
\end{figure}

Figure \ref{fig:effects_simple} shows average effects of the most important determinants of transmission. Age-specific effects for all covariates are available in Supplementary Figure \ref{fig:effects}. Interventions change the reproductive number by a fixed ratio. Interaction effects are ignored and estimates evaluate the effect of an intervention under the circumstances it was implemented.

Many covariates are strongly correlated (Figure \ref{fig:cor_main}). However, the correlation matrix of effect estimates suggests that effects are identified (Figure \ref{fig:estimates_corr}).

The strongest effect was associated with \emph{public awareness rising}. Transmission reduced by $58\%$ ($95\%$-CI: $[53\%,62\%]$) when government officials gave their first speeches asking for decided behavioral changes from the public. Simultaneously implemented changes, like the staggered declaration of international risk areas, might bias this result.

The \emph{closure of restaurants} was associated with additional reductions in transmission of $15\%$ ($95\%$-CI: $[5\%,23\%]$). The effect was strongest for the age group 15-59.

\emph{School and daycare closures} were associated with a $12\%$ ($95\%$-CI: $[-5\%,28\%]$) reduction in transmission. The intervention affected the age group 15-79 equally, which suggests that the reduction stems from behavioral changes, instead of the absence of transmission in the educational setting, which would affect other age groups one generation time later.

\emph{Limiting sports} was associated with additional reductions in transmission. The \emph{closing of non-essential shops}, mandatory \emph{distancing} and limitation of \emph{gatherings} in public spaces, however, showed no evidence for reducing transmission. In the age group 15-34, transmission actually \emph{increased} by  $25\%$ ($95\%$-CI: $[-1\%,54\%]$) when public distancing became mandatory. This might explain why no significant effect of stay-at-home orders nor of business closure was found in the United States \cite{CHERNOZHUKOV2020}. The regulation could induce more private contacts with higher transmission risk. In some states a mild \emph{stay-at-home order} was imposed, allowing individual sport and work, which was associated with a  $9\%$ ($95\%$-CI: $[4\%,13\%]$) reduction in transmission across all age groups.

Initially, testing was covered for patients with exposure or within a risk group. Later, any symptomatic case, regardless of exposure, was eligible. The \emph{narrow testing} regime was associated with a  $16\%$ ($95\%$-CI: $[7\%,26\%]$) increase in transmission. This effect is driven by younger cohorts, whereas transmission in the age group 80+ was reduced.

The \emph{reopening of daycare and churches} with precautionary measures had no significant impact. Some evidence for increased transmissions was associated with school reopenings.

\emph{Masks} (mouth and nose cover including cloth masks) became mandatory in supermarkets and public transport. Compliance was found to be high \cite{betsch2020social} and timing of implementation varied across states and counties. A small effect of $-6\%$ ($95\%$-CI: $[-17\%,7\%]$) was associated with this policy. If only a small ratio of transmissions occurs in these settings, this is consistent with a household study in China, where mask wearing before symptom onset reduced transmission by $80\%$. As a comparison, mandatory masks for employees were associated with a reduction in transmission by $10\%$ in the United States \cite{CHERNOZHUKOV2020}.

\subsubsection*{Effects of information and testing and tracing}

Testing and tracing has been argued to be crucial \cite{ferretti2020quantifying}. As symptom onset date and reporting date (when health departments were informed about positive test) are available, we know for each case the days of potential infectiousness and when testing and tracing was initiated (for details see the supplementary material). This allows to construct a daily location specific proxy for the \emph{ratio of traced infectious} and to provide empirical evidence on the effectiveness of testing and tracing that corroborates results of modelling studies \cite{kretzschmar2020impact}.

The ratio of traced infectious was associated with a reduction of transmission by $33\%$ ($95\%$-CI: $[22\%, 43\%]$). As the ratio does not reflect unreported infections, the \emph{impact of testing and tracing on an infectious individual} arises after adjusting for the reporting rate. Extrapolating to unobserved infections, testing and tracing reduced secondary infections by $84\%$ ($95\%$-CI: $[35\%, 100\%]$) for the age group 15-59. The effect was strongest for the high risk group over 60 years, where the ratio of traced infectious reduced transmission by $44\%$ ($95\%$-CI: $[27\%, 59\%]$). Adjusting for unreported infections, this effect corresponds to an eradication of transmission to older age groups for tested and traced infectious individuals. The hypothetical extrapolation to all infections should be interpreted with caution.

\begin{figure}[htp]
	\centering
	\includegraphics[width=.8\linewidth]{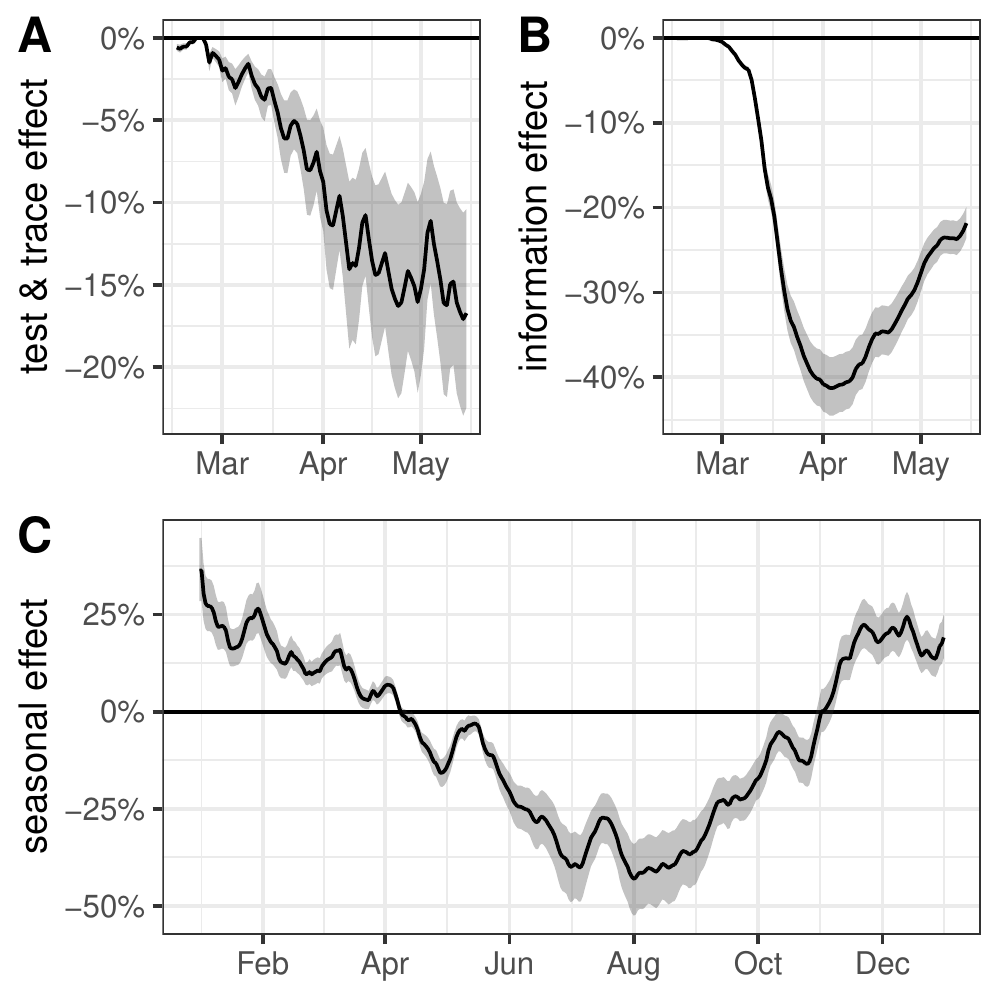}
	\caption{Total effect of testing and tracing (ratio of traced infectious), information (logarithm of reported local incidence), and season (average temperature and relative humidity). $95\%$ confidence bands are shown. Figure A and B denote total effect given the data. Figure C \emph{extrapolates} the total effects of weather variables in an out-of-sample prediction based on average daily weather in the past three years, where confidence bands represent uncertainty in effect estimation, and results are smoothed with a 14-day rolling average.}
	\label{fig:total_all}
\end{figure}

\maybe{The effect of testing and tracing conflates at least two potential channels. First, individuals adapt their behavior after confirmed infections. Second, the local health department administers quarantine and traces contacts.}

Recent modelling studies for Germany concluded that 30-50$\%$ of transmission can be reduced by testing and tracing \cite{contreras2020challenges}. I find that the total effect of testing and tracing increased from March to May (Panel A of Figure \ref{fig:total_all}). In May 2020, testing and tracing was attributed an average transmission reduction of $15\%$ ($95\%$-CI: $[9\%,20\%]$). For the age group over 80 years, the impact is larger with a reduction of $28\%$ ($95\%$-CI: $[17\%,37\%]$).

In general testing and tracing capabilities are limited. If infections surpass capacities, the ratio of traced infectious decreases. The results from above imply that unmitigated spread increases the speed of transmission, especially in older age groups.

Another potential factor for transmission is voluntary behavioral response to risk of infection \cite{CHERNOZHUKOV2020}. Individual risk of infection depends on local \emph{incidence}. In line with this argument, I find that publicly reported incidence reduced transmission. This \emph{information effect} was highest in April (Panel B in Figure \ref{fig:total_all}) when estimates indicate a reduction of  $44\%$ ($95\%$-CI: [$40\%$, $48\%$]).

An interesting counterfactual is the level of incidence that is sufficient to stop growth in the absence of policy interventions. The results suggest that a reported incidence of 300 to 1000 weekly cases in 100,000 induces sufficient voluntary behavioral change. This incidence corresponds to 3,600 to 12,000 weekly deaths in Germany (based on a reporting rate of $50\%$ and an infection fatality rate of $0.75\%$ guided by \cite{levin2020assessing} and the German age distribution from 1 to 80 years). The result is speculative, as it extrapolates beyond the support of the data (the $99\%$ quantile of incidence is 160 cases in 100,000).

\emph{Cumulative incidence} (in percentage points) was negatively associated with transmission. Random mixing, full immunity after infection, and the absence of underreporting would be consistent with an effect of $-1\%$. The data does not allow a sharp identification of this effect as the highest measured cumulative incidence was $1.2\%$, which corresponds to a reduction of $20\%$ ($95\%$-CI: $[4\%,35\%]$). \maybe{The effect is strongest for older age groups (Figure \ref{fig:effects}). A possible explanation for the age dependence would be that younger population groups are more likely to react to recent reports on incidence, whereas older age groups react slower to current news but behavioral changes are more persistent.}

\subsubsection*{Effects of weather and seasonality}

Seasonal effects have been discussed to play a role in SARS-CoV-2 spread \cite{baker2020susceptible,carlson2020misconceptions}. I find that low temperature was associated with higher transmission. Relative humidity was associated with a small increase in transmission. As the latter effect is mostly driven by the age group 60-79, there is no strong evidence for relative humidity being an important determinant of transmission in Germany. The extrapolated change in transmission due to seasonality is $53\%$ ($95\%$-CI: $[43\%, 64\%]$) (based on average temperature and relative humidity in January and July). The expected seasonal effects over the year are denoted in Panel C of Figure \ref{fig:total_all}. These findings are consistent with other studies. In an international cross-city comparison, low temperature and humidity was associated with community spread \cite{sajadi2020temperature}. A cross state comparison in the United States found that high temperature and UV-light reduced incidence \cite{sehra2020maximum}.

The effect estimate for seasonality conflates the interaction of the virus with environmental circumstances and behavioral changes due to weather \cite{carlson2020misconceptions}. Moreover, extrapolating effects identified by daily weather to entire seasons can induce substantial biases. \maybe{Behavioural changes might be driven by relative changes to recent weather, instead of absolute values, which would lead to an overestimation of seasonal effects. Other effects of weather on transmission might accumulate, which would lead to an underestimation of seasonal effects.}In line with the findings presented here, a study based on monthly case data in countries on both hemispheres found that the season of other human corona viruses was associated with low temperature and high relative humidity \cite{Li2020seasonality}.

\subsubsection*{Recent data as robustness check}

The effect estimates presented so far are subject to substantial uncertainty and suffer from high correlation (compare Figure \ref{fig:estimates_corr}). Model misspecification may play a role as other factors can influence behavior contemporaneously. 

The main model is estimated on data until May 2020 as detailed intervention data is not available for the remaining time period. However, the covariates for weather, information, and testing and tracing are available. As robustness check, I consider data from May to August 2020 including the age group 5-14 and with the same model specifications, but substituting policy interventions with week fixed effects. This study covers another 30,000 cases in 141 counties. See the supplementary material for details.

I find that results are consistent for information on incidence, temperature, and testing and tracing. Seasonality effects were less strong which might indicate non-linear effects. School children below 15 years exhibited higher transmission during school holidays. This suggests that reopened schools constituted a relatively low risk of transmission.

\subsubsection*{Discussion}

Empirical evidence is paramount to inform modelling studies and policy decisions. I rely on features of German surveillance data to improve upon early empirical studies analyzing SARS-CoV-2 transmission \cite{flaxman2020estimating,hsiang2020effect,salje2020estimating,dehning2020inferring}. In particular, symptom onset allows sharper estimation of infection timing. Moreover, age groups improve identification of growth rates and enable the estimation of age-dependent effects. Inference in such highly compartmentalized data should account for superspreading. As with any structural estimation, results should be interpreted with caution and a keen eye on the model's limitations. The Bayesian framework allows to integrate alternative assumptions (e.g., priors on the reporting rate) to asses their impact on the results presented here.

A large set of covariates was analyzed. Previous studies found stronger effects of policy interventions without controlling for information, tracing, or seasonality. For example, the shut-down in France was attributed a $77\%$ reduction in transmission based on hospitalization and death data \cite{salje2020estimating}. An international comparison of death data estimated an effect of $81\%$ \cite{flaxman2020estimating}. A reduced form approach on case growth found heterogeneous effects of policy interventions in China, South Korea, Italy, Iran, France and the United States \cite{hsiang2020effect}. As noted in other studies, growth rates in Germany substantially reduced before major policy interventions were put in place \cite{dehning2020inferring}. If not controlled for, adaptation to local risk of infection can be wrongly attributed to policy interventions \cite{CHERNOZHUKOV2020}.

\maybe{Contact tracing data is an alternative approach to provide empirical evidence. While contact tracing data is prone to oversampling close contacts and large clusters, incidence data (reported cases or deaths) is more likely to detect seemingly unrelated cases and large sample sizes are available, at the cost of additional modelling assumptions. Neither of the two methods is strictly dominated in terms of reliability by the other.} 

\maybe{The model presented here can be applied in a similar manner to death counts or cases by reporting date by interchanging the incubation period distribution with the distribution from infection to death or reporting. As those are more dispersed, inference would deliver less sharp estimation. This illustrates the importance of aggregated cased data with symptom onset.}

I find that higher reported incidence reduced transmission. This explains the common observation that incidence rarely exhibits prolonged exponential growth. Spread is unmitigated as long as it is undetected. In Germany, the reduction in transmission was driven mainly by \emph{behavioural adaptations} to reported incidence instead of \emph{immunity} by infection. If transmission keeps decreasing with higher incidence, there exists an equilibrium value for \emph{reported incidence} where the reproductive number is $1$. If information or behavioral adaptation is delayed, incidence moves in waves around this equilibrium value.

\maybe{It can be argued that sensibility to local incidence is changing over time. On the one hand constant exposure to the risk might reduce perceived risk and therefore behavioral adaptions. On the other hand experience could improve the effect of individual prevention efforts.}

Individual risk of contracting (or spreading) Covid-19 is a function of behavior and local incidence. If behavior is constant, incidence grows until immunity slows further expansion. Behavioural changes to prevent transmission are costly to individuals and their peers, but decrease incidence and are therefore potentially beneficial to society. Thus, policy interventions can be warranted to share and direct the costs of transmission prevention. As behavior was found to change with local incidence, indirect social and economic consequences of the pandemic can be expected even in the absence of policy interventions. Ultimately the impact of policy interventions depends on their effect on transmission \emph{and} their ability to allocate costs of prevention effectively.

Estimating the costs of interventions is beyond the scope of this study, but some remarks regarding effectiveness seem warranted: Public awareness rising was crucial, while little effect of closing shops and public distancing rules was found. Daycare reopening showed no, school reopening relatively little impact. Testing and tracing had a large effect. As many infections occur before or around symptom onset, timely testing and fast turnaround is important. These results support the call for investments in alternative testing technologies \cite{mina2020rethinking}. 

Seasonality of SARS-CoV-2 transmission has been a highly politicized topic \cite{carlson2020misconceptions}. Given the high rate of susceptibility, seasonal effects on SARS-CoV-2 kinetics are limited \cite{baker2020susceptible}. While weather was only one of the key determinants of transmission in the study period, its extrapolated effect could be decisive. Importantly, seasonality increases the equilibrium incidence where information stalls growth. As testing and tracing capacities are limited and crucial to prevent transmission (especially to older age groups), higher incidence could lead to an additional increase in growth rates and an unproportional increase of exposure in high risk groups.

\bibliographystyle{plain}

\bibliography{scibib}

\section*{Acknowledgments}
For helpful discussions and comments I thank Johannes Bracher, Holger Brandt, Timo Dimitriadis, Simon Heß, and Zachary Roman. Yalini Ahrumukam provided excellent research assistance.\\
Funding: The work of Patrick Schmidt has been funded by the SNF.\\
Competing interests: The author declares that he has no competing interests. \\
Data and materials availability: Data was obtained from the Robert Koch Institute and the German Weather Service. All data is available in the manuscript or the supplementary materials. The data and the replication code are available at \url{https://github.com/Schmidtpk}.

\makeatletter\@input{xx2.tex}\makeatother
\end{document}


\maketitle

\tableofcontents

\section{Model}\label{sec:supp_model}

\begin{figure}	
	\centering
	\small
	\begin{tikzcd}[cells={nodes={draw=gray, circle}}] 
	\dots 	& i_{t-2} \arrow[rrr, "D_i(2)R_t" description, bend left, shift left=1] 
	& i_{t-1} \arrow[rr, "D_i(1)R_t" description, bend left] 
	&
	& |[draw=black,thick,fill=lightgray]| i_t 
	\arrow[rrd, "D_s(1)r_{t+1}\qquad\quad" description,bend left = 5, shift right=2] 
	\arrow[rrrd, "D_s(2)r_{t+2}" description,bend left = 5, shift right=2] 
	\arrow[rrr, "D_i(2) R_{t+2}" description, bend left, shift left=1] 
	\arrow[rr, "D_i(1) R_{t+1}" description, bend left]
	& & i_{t+1}   &  i_{t+2}          &   \dots           \\
	&   &  \dots  &  & c_t  &    & c_{t+1} & c_{t+2} &  \dots\\
	\end{tikzcd}
	\caption{Model illustration. All nodes directly connected to new infections $i_t$ are shown. Each infection $i_{t'}$ with $t'<t$ transmits on average to $D_i(t-t')R_t$ new cases, where $D_i$ denotes the generation time distribution. Infections at time $t$ develop symptoms at time $t'>t$ with probability $D_s(t'-t)$ and are reported with probability $r_{t'}$.}
	\label{fig:model}
\end{figure}
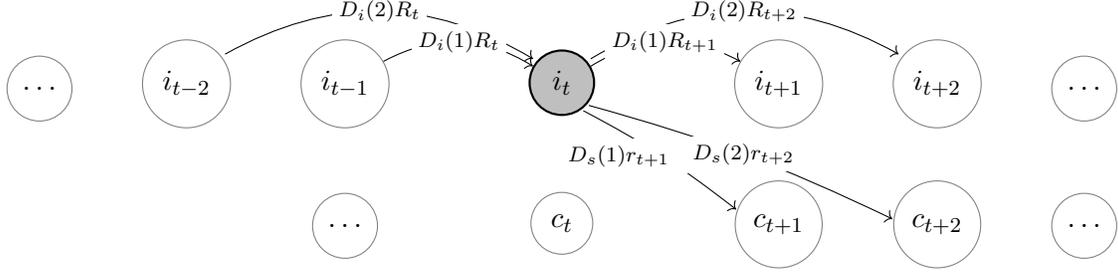

The model comprises three parts. The transmission model denotes the transmission dynamic over time. It accounts for dispersion in secondary transmissions and a probabilistic latent period, but does not include importation (across location or age). The key feature is a time dependent and age-specific instantaneous reproductive number. Infections are unobserved, and have a probabilistic connection to reported cases in the measurement model. The measurement model features a probabilistic incubation period and a probability of symptom development being detected and reported. The effect model describes how covariates (e.g., interventions, weather, information, etc.) influence the instantaneous reproductive number. All other characteristics (e.g. dispersion, latent period distribution, etc.) are assumed to be constant.

\subsection{Transmission}

The number of infections in location $l$ and age group $a$ is denoted by $i_t^{l,a}$. Infections $i_t^{l,a}$ lead to additional infections by transmission. The transmission model has two key features: Firstly, the current average growth rate is represented in the instantaneous reproductive number $R_t^{l,a}$. Secondly, the model allows for dispersion (compare \cite{lloyd2005superspreading}) of secondary infections, as overdispersion is hypothesized to be a crucial component of Covid-19 infection dynamics \cite{endo2020estimating,adam2020clustering}.

Let $i_{t,t'}$ denote the number of infections at time $t$ caused by primary cases infected at time $t'$, where we omit location and age for notational convenience. We have $i_t = \sum i_{t,t'}$.
Let $i_{j,t,t'}$ for $j=1,\dots,i_{t'}$ denote the offspring at time $t$ of individual $j$ infected at time $t'$. Let $i_{t',j} = \sum_t i_{j,t,t'}$ denote the sum of secondary cases of individual $j$. All distributional statements for random variables realizing at time $t$ are meant given $R_t$ and previous infections $i_{t-1},i_{t-2},\dots$ if not noted otherwise. Let $D_i$ denote some constant generation time distribution with positive support \cite{cori2013new}.\\

The negative binomial distribution of secondary infections can be motivated as a generalization of the Poisson model (as mixture of Poisson distributions, where the mixing distribution of the Poisson rate is gamma distributed \cite{lloyd2005superspreading}), or as the mechanistic outcome of a linear growth process \cite{finkenstadt2000time}.

Assumption:\\ The offspring $i_{j,t,t'}$ has a negative binomial distribution with mean $R_t D_i(t-t')$ and dispersion $\Psi D_i(t-t')$ and is independent for $t'<t$ and $j \in \{1,\dots,i_{t'}\}$.\\

It follows that $i_t \sim NB(R_t L_t, \Psi L_t)$, where $L_t = \sum_{t'} D_i(t-t')i_{t'}$ denotes the viral load circulating at time $t$. The argument is as follows: As $i_{j,t,t'}$ for $j=1,\dots,i_{t'}$ are independent and identically distributed, it follows that $$i_{t,t'}=\sum^{i_{t'}}_{j=1} i_{j,t,t'} \sim NB(R_t D_i(t-t') i_{t'}, \Psi D_i(t-t') i_{t'}).$$
As $i_{t,t'}$ for $t'<t$ are independent and have a negative binomial distribution with identical parameter $p_{t,t'}=\frac{\Psi}{R_t+\Psi}$, the distribution of the sum of all transmissions at day $t$ is denoted by
$$i_t = \sum_{t'<t} i_{t,t'} \sim NB(R_t L_t, \Psi L_t).$$

Note the difference to the widely used parametrization of dispersion in \cite{lloyd2005superspreading}, which models the dispersion of the amount of secondary infections without generation time. If the reproductive number is constant, i.e. $R_t=R$, a single infection induces $\sum_t i_{j,t,t'} \sim NB(R,\Psi)$ secondary infections, coinciding with the aforementioned standard model of dispersion in the seminal paper \cite{lloyd2005superspreading}. Thus, priors on the dispersion $\Psi$ can be informed by studies working in the standard framework and posterior results can be compared conveniently. Notably, if the instantaneous reproductive number is varying over time, the number of secondary infections constitutes a \emph{mixture} negative binomial distribution instead as denoted in \cite{furman2007convolution}.

\maybe{The dispersion model above allows to estimate a valid and easily interpretable dispersion parameter from daily case reports. Previous studies had to rely on contact tracing data \cite{lloyd2005superspreading} or multiple transmission chains/outbreaks \cite{blumberg2013inference}.}

If $D_i(1)=1$ (a common assumption for weekly case counts), the arguments presented here simplify. Specifically, the viral load is $L_t=i_{t-1}$ and the assumption reduces to $i_{j,t,t-1}$ being independent given $R_t$ and having a negative binomial distribution with mean $R_t$ and dispersion $\Psi$. We note the similarity to \cite{wakefield2019spatio} who derive a negative binomial distributions for the time series SIR model first introduced in \cite{finkenstadt2000time} from a linear birth process, considering only a single time step, in which case the model presented here would reduce to $i_t \sim NB(R_t i_{t-1}, i_{t-1} \Psi)$, which recovers the time series SIR model for $\Psi=1$. Thus, the model presented here can be seen as a generalization of the time series SIR model with flexible dispersion $\Psi$ and accounting for a generation distribution $D_i$.

Consider the alternative assumption generating standard models from the literature:\\ The offspring $i_{j,t,t'}$ has a negative binomial distribution with mean $R_t D_i(t-t')$ and dispersion $\Psi D_i(t-t')$ and $i_{j,t,t'}=i_{j',t,t'}$ for all $j,j' \in \{1,\dots,i_{t'}\}$.\\

It follows that $$i_{t,t'}=\sum^{i_{t'}}_{j=1} i_{j,t,t'} = i_{t'} i_{j,t,t'} \sim NB(R_t D_i(t-t') i_{t'}, \Psi D_i(t-t')).$$
As $i_{t,t'}$ for $t'<t$ are independent and have a negative binomial distribution with identical parameter $p_{t,t'}=\frac{\Psi}{R_t+\Psi}$, the distribution of the sum of all transmissions at day $t$ is denoted by
$$i_t = \sum_{t'<t} i_{t,t'} \sim NB(R_t L_t, \Psi).$$ If transmission is considered only between subsequent time intervals, we obtain the standard assumption of inference based on negative binomial regression, the endemic/epidemic model introduced in \cite{held2005statistical}, or epidemiological models with random effects (e.g., the model in \cite{fisher2020ecological}), i.e. $i_t \sim NB(R_t i_{t-1}, \Psi)$. \maybe{Arguably, this alternative assumption that all transmissions induced by different individuals infected at time $t'$ are identical given the instantaneous reproductive number $R_t$ stands in contrast to the goal of modelling the average variation in reproductive.} A similar point applies to \cite{flaxman2020estimating}, where dispersion is independent of the total count in the death reports.

\subsection{Measurement}

As illustrated in Figure \ref{fig:model}, it is assumed that an infection at time $t$ leads to symptom onset being reported at time $t'$ with probability $D_s(t'-t)r_t$, where $D_s(t'-t)$ denotes the distribution of the incubation period and $r_t$ the likelihood of developing symptoms \emph{and} being positively tested and reported as a case. The aggregation of all cases $c_t$ constitutes a sum of Bernoulli trials, following a Poisson binomial distribution, which is approximated by a Poisson distribution for computational convenience: 
$$c_t \sim Poisson(r_t \sum_{t'<t} i_{t'} D_s(t-t')).$$

Extensions of the model to include additional observable case counts are straight forward. The paper focuses on symptom onset of Covid-19 cases and argues that this provides sharper information on the  timing of infections than death counts or hospital admissions. Deaths counts would require that the infectuous fatality rate is constant or adequately modelled. Hospital admissions and deaths counts suffer additionally from lower numbers, especially among younger age groups. Aggregated reported cases (without symptoms) include asymptomatic cases, but require knowledge on the timing from infection to reporting and are subject to larger distortions if proportion of asymptomatic infections ascertained as cases changes due to testing regime.

\subsection{Effects}
The instantaneous reproductive number is modelled as a function of input covariates $x_1,\dots,x_J$, where the effect is assumed to be multiplicative such that

$$R_t^{l,a} = R_0^{l,a} \prod_{j=1}^J (1 + \beta^a_j x_j).$$

Each location and age group has an individual basic reproductive number $R^{l,a}_0$. The effect of covariate $x_j$ on the instantaneous reproductive number of age group $a$ is denoted by $\beta^a_j$. Covariates are usually standardized or dummies. A coefficient of $.5$ would signify that an increase from the 0 to 1 in $x_j$ increases the instantaneous reproductive number by $50\%$. A coefficient of $-0.5$ would signify that the same change in $x_j$ is associated with a $50\%$ reduction in transmission.

\section{Data}\label{supp:data}

The study combines information from three main sources. Case reports are obtained by the Robert Koch Institute (RKI)\footnote{\url{https://npgeo-corona-npgeo-de.hub.arcgis.com/datasets/dd4580c810204019a7b8eb3e0b329dd6_0}}, weather data from the German weather service\footnote{Accessed with the R-Package rdwd \cite{rdwd}.}, and policy interventions were specifically catalogued for this study\footnote{Source document with footnotes accessible at \url{https://docs.google.com/spreadsheets/d/1cmGBMUhBt5y6jwiqaF7lh7VQNMN6D5FCoIltlOzhMfI/edit#gid=0}}. The data is available in an accompanying R-package.\footnote{Available at \url{https://github.com/Schmidtpk/CovidGer}.}

\subsection{Cases}

Germany recorded over 170,000 cases until 15 of May 2020 of whom 73\% reported a date of symptom onset. The first case was reported in January, but major outbreaks started only end of February. It has been argued that Germany had relatively large testing capacities in the early phase of the pandemic \cite{wieler2020emerging}. This is also supported by excess death data \cite{stang2020excess}, which provides no evidence for undetected Covid related deaths.

The data was obtained by the RKI, the research institute responsible for disease control and prevention that is subordinate to the Federal Ministry of Health in Germany. Laboratories are required by law to report positive test results within 24 hours. The data is gathered by local health departments (Gesunheitsamt) responsible for collecting reports on \emph{notifiable diseases} (diseases required by law to be reported to government authorities) and subsequently passed along to the RKI. The health departments are organized on a subregional level (county or ``Landkreis'').

The data includes case-specific information on date of symptom onset, date of reporting to the health department, age bracket, county, and death. More detailed information (occupation, likely transmission environment,etc.) is provided to the RKI, but not publicly available \cite{IfSG}.

\begin{figure}
	\centering
	\includegraphics[width=.8\linewidth]{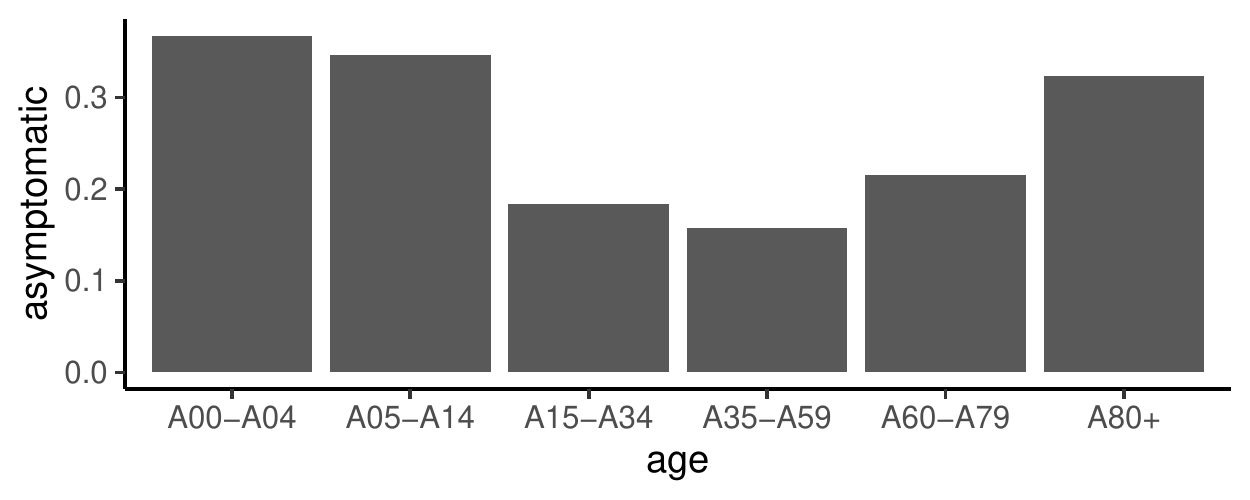}
	\caption{Ratio of asymptomatic cases by age group. Ratio of cases reported without symptom onset. This may include asymptomatic cases, presymptomatic cases that were not followed up on, or symptomatic cases, where symptom onset was not reported.}
	\label{fig:asymptomatic}
\end{figure}

Not all cases reported a date of symptom onset. Generally, entries are updated after first date of reporting, which suggests that symptom onset after testing is reported. The remaining cases without symptoms are either asymptomatic cases, false-positive tests, or cases were symptoms were developed but not consistently reported. Interestingly, the ratio of asymptomatic cases is age-dependent as illustrated in Figure \ref{fig:asymptomatic}. This suggests different likelihood of developing symptoms across age groups, with the age group 35 to 59 years being most likely to exhibit symptoms. An age-dependent likelihood of developing symptoms provides another argument for the relevance of modelling the growth of Covid-19 with age compartments to disentangle a changing age distribution and changing incidence. 

\begin{figure}
	\includegraphics[width=\linewidth]{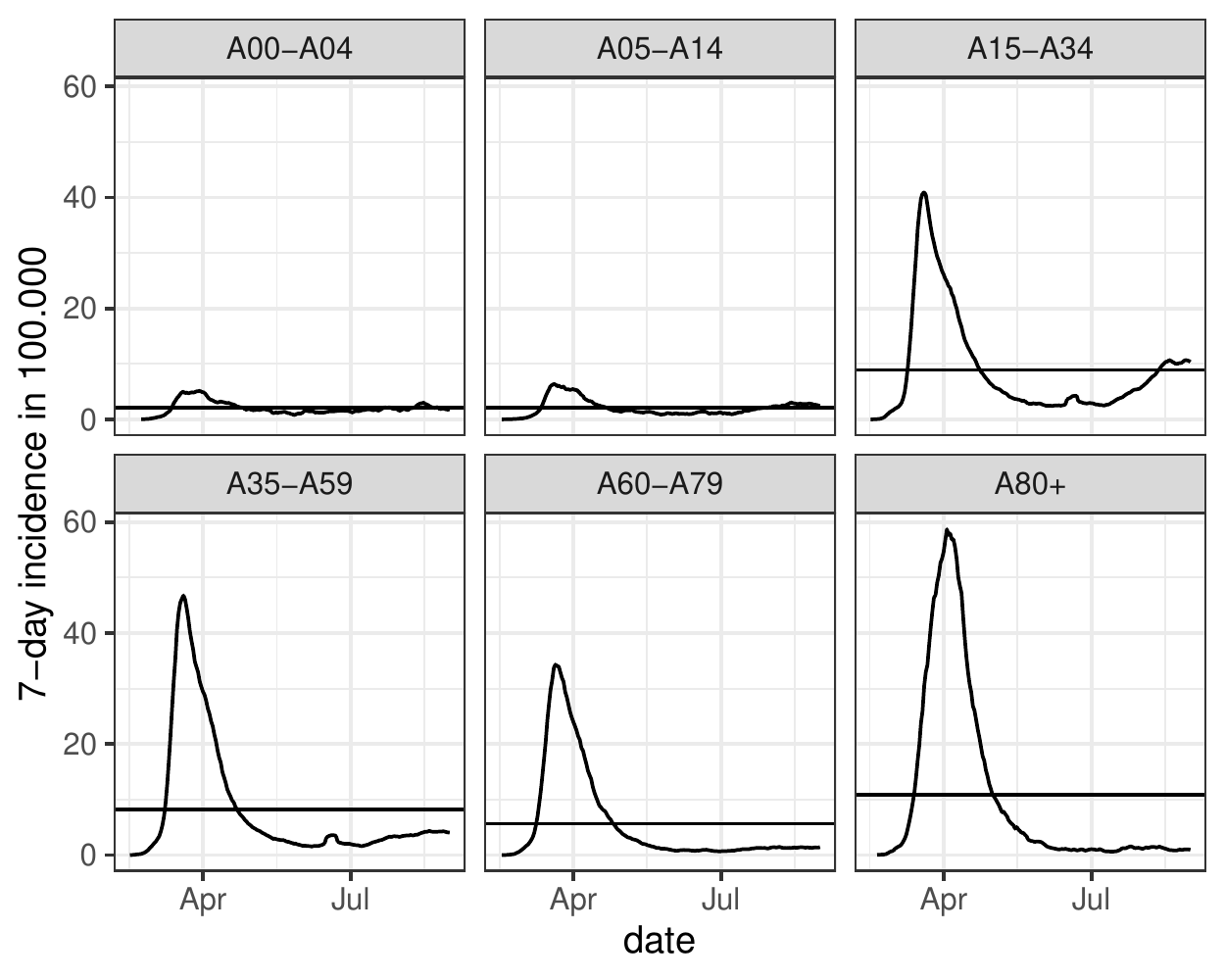}
	\caption{7-day incidence by age group. Horizontal lines denote age-specific average incidence. Cases are based on symptom onset. Asymptomatic cases are not included.}
	\label{fig:incidence}
\end{figure}

Figure \ref{fig:incidence} shows the 7-day incidence of symptomatic cases across age groups\footnote{Population data was obtained from the Regional Database \cite{regionaldatenbank}.}. Notably, the reported cases were first dominated by moderate age groups, in April overtaken by the elderly, and in the recent past mainly driven by the age group 15 to 34 years.

Note that the age groups 0-4 and 5-14 were not included in the main model as relatively few cases were observe in the study period until May 2020. Further, the low infectious fatality rate makes an assessment of the reporting rate by the case fatality rate (see Figure \ref{fig:cfr_main} in the main text) more challenging. A large-scale serological study in the state of Bavaria (Bayern) found that children were 6 times more likely to be sero-positive than expected by case reports \cite{hippich2020public}, which suggests that their reporting rate was about half as large as those of adults.

\begin{figure}
	\includegraphics[width=\linewidth]{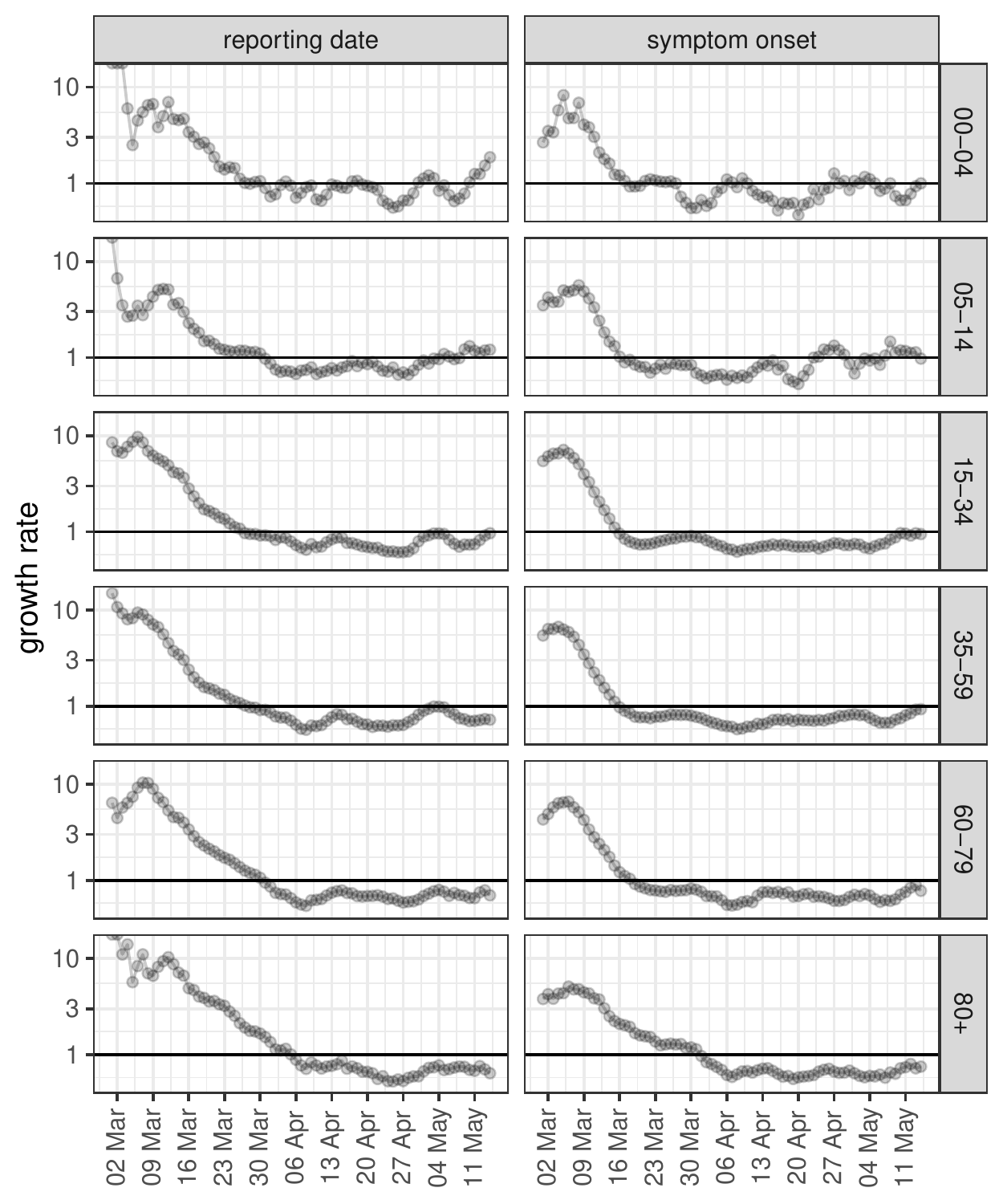}
	\caption{Weekly growth rate of cases. Growth rate of 7 day rolling window of cases in study period. Reporting dates are shifted by 10 days, symptom onset by 5 days.}
	\label{fig:growth}
\end{figure}

Figure \ref{fig:growth} shows growth rates computed on a 7-day window to account for weekday effects. The left column is based on reporting dates. The right column based on symptom onset.

\subsubsection{Outbreak in Heinsberg}

The first major outbreak recorded was in the community Gangelt in the county Heinsberg \cite{streeck2020gangelt}. Indeed, 47\% of the 734 cases that were officially recorded with symptom onset in February 2020 were reported in this county.

\begin{figure}
	\includegraphics[width=\linewidth]{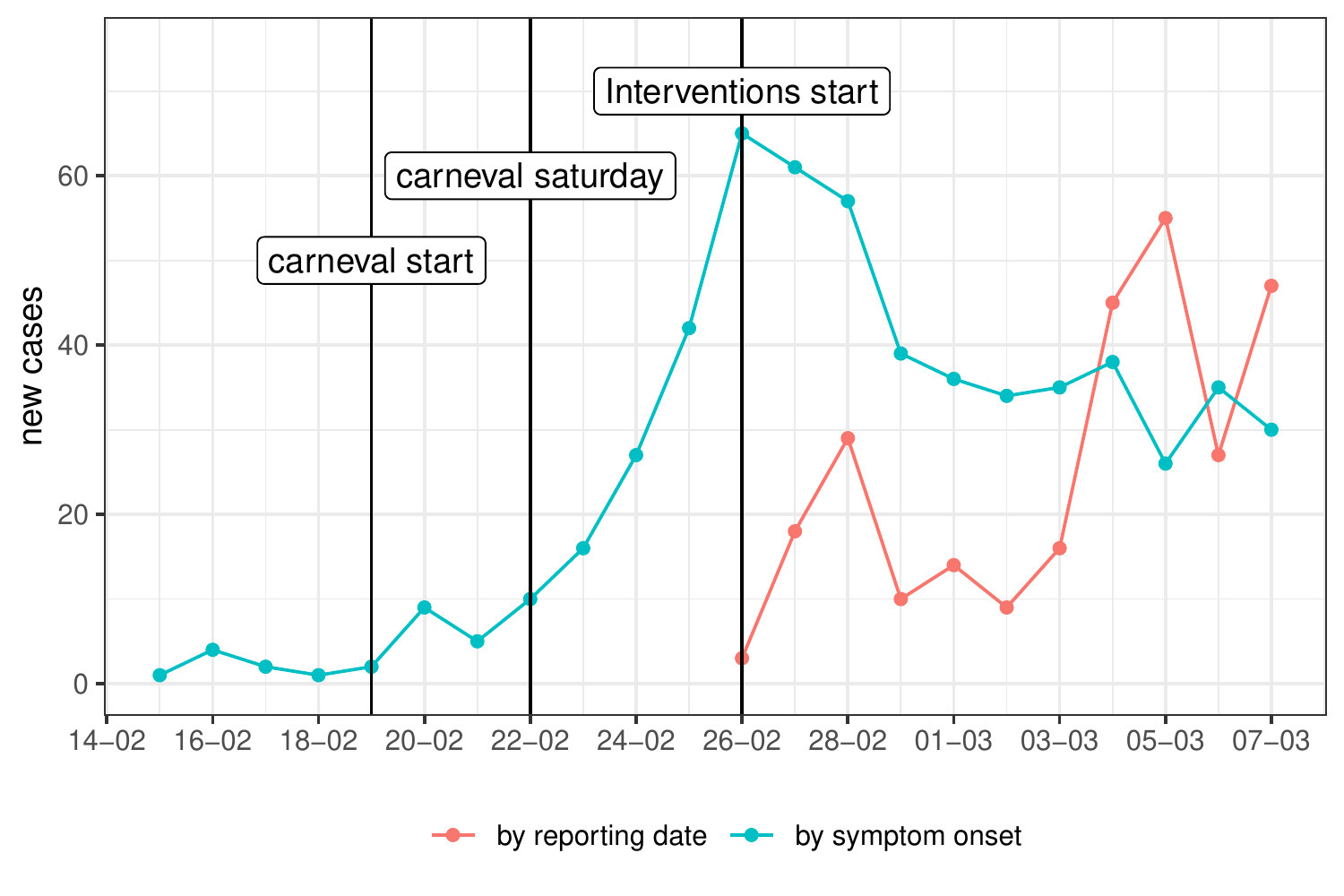}
	\caption{Daily new cases in county Heinsberg by symptom onset and reporting date.}
	\label{fig:heins}
\end{figure}

Figure \ref{fig:heins} illustrates the advantage of using symptom onset to judge the timing of infections. We observe a clear pattern of constant growth rates of new cases with symptom onset until February 26. With an incubation period of 4-7 days this suggests that reproductive numbers were high during carneval, which is celebrated most intensely between the Thursday (February 19) and Saturday (February 22). At February 26, local authorities became aware of the first cases and closed schools and daycares. In the following days a large number of people quarantined. The figure illustrates that reporting date shows a significant lag and is not appropriate to connect infections to specific circumstances.

As the situation in Heinsberg was extraordinary, with a large media exposure and drastic adaptations beyond the officially recorded measures, the county is excluded from the main analysis.

\subsection{The variance of growth rates}\label{supp:variance_growth}

In the model section of the main text, where generation time is assumed be $1$, it is argued that the realized growth rate 
$g_t = \frac{i_t}{i_{t-1}}$ has a variance of $\sigma^2_{g_t}=\frac{R_t (\Psi+R_t)}{i_{t-1}\Psi}$ for overdispersed individual transmission. 

\begin{figure}[htp]
	\centering
	\includegraphics[width=\linewidth]{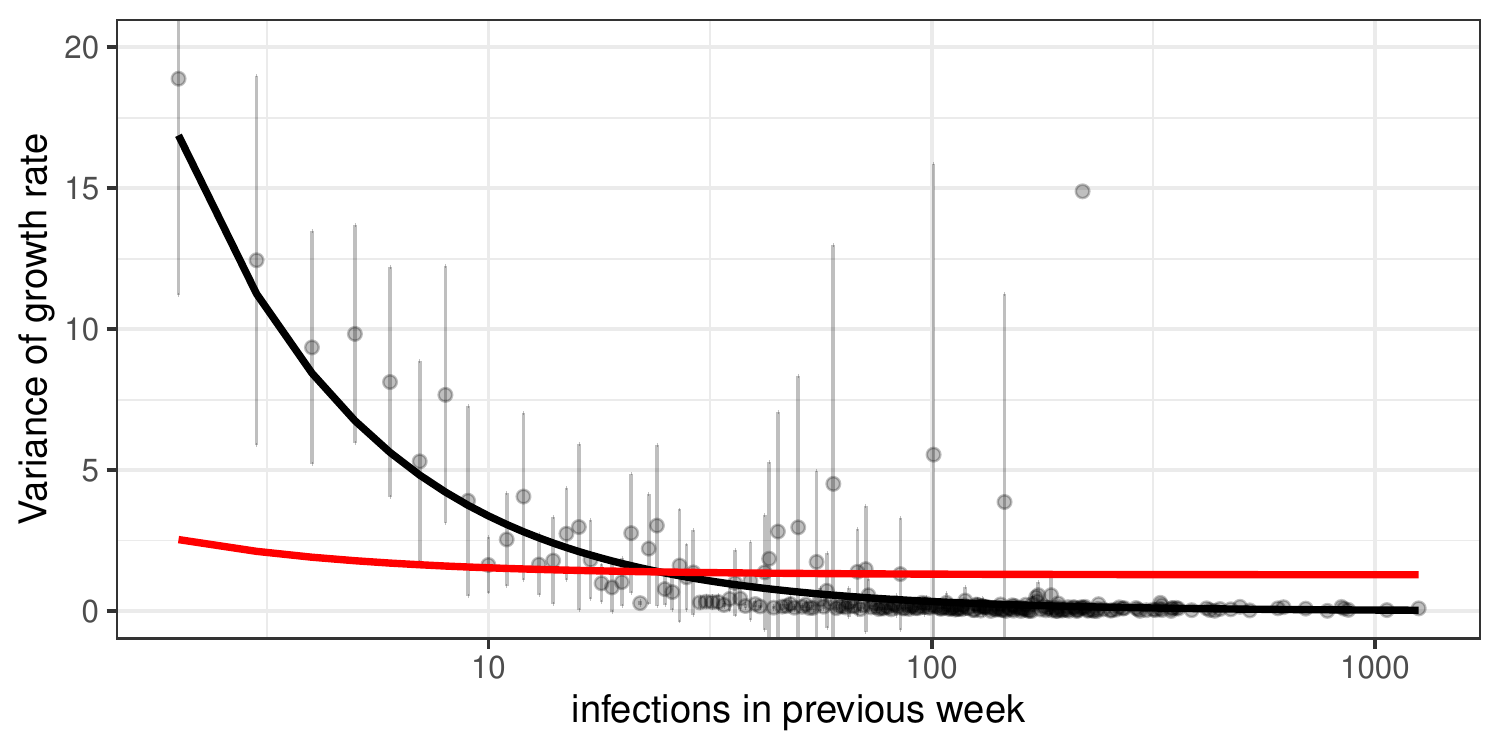}
	\caption{Empirical variance estimates with $95\%$-confidence intervals based on normal assumption and estimated variance. The black line denotes predicted variance of growth rates based on $\sigma^2_{g_t}=\frac{R_t (\Psi+R_t)}{i_{t-1}\Psi}$ with $R=2.5$ and $\Psi=0.2$. The red line denotes predicted variance based on the alternative assumption as denoted in the main text.}
	\label{fig:grate_ilag}
\end{figure}

This point is investigated in the data. Weekly symptomatic cases were computed such that the assumption of a generation time and incubation period of $1$ seems reasonable and weekday effects are accounted for.
First, we estimate the variance for each value of $i_{t-1}$ in the data by its empirical analogue. Figure \ref{fig:grate_ilag} illustrates variance estimates and their fit with the modelled derived in the main text.

The finding can also be used to estimate the dispersion over time. In particular, it follows directly that dispersion is a function of the variance of growth rates:
$$\Psi = \frac{R_t^2}{\sigma_{g_t}i_{t-1}-R_t}.$$
We evaluate the development of dispersion over time using the equation above. $R_t$ is assumed to be constant within a state and estimated by the growth rate average across counties. As estimator for $\sigma^2_{g_t}$ we use its sample analogue within state and month.

\begin{figure}[htp]
	\centering
	\includegraphics[width=\linewidth]{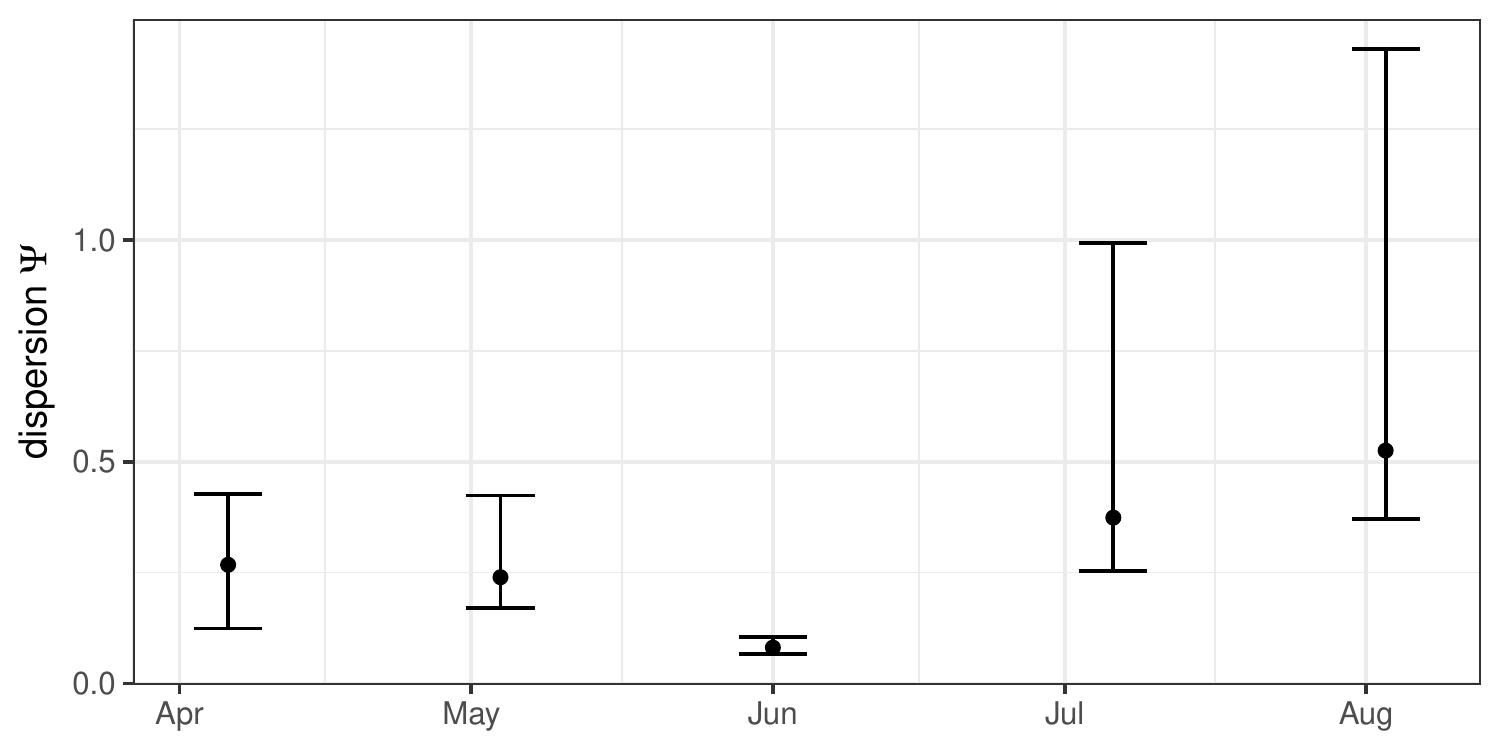}
	\caption{Reduced form estimation of dispersion with $95\%$-confidence intervals based on variance estimation only.}
	\label{fig:disp_reduced}
\end{figure}

Basically, it is assumed that average number of secondary transmissions is constant within state, and the variance of growth rates within a state across counties is used to estimate the dispersion parameter. Results can be seen in Figure \ref{fig:disp_reduced}. Without accounting for specific effects of interventions and covariates, the dispersion $\Psi$ is estimated to be around $0.25$ in the first months, lower in June and increasing in summer. Those estimates are slightly below the one obtained in the full model, where the average across all age groups for cases until mid May is estimated to be consistent with a dispersion parameter of $0.47$.

\subsection{Cumulative incidence}
Several explanatory variables were constructed based on the case data provided by the RKI. This includes the \textit{cumulative incidence}, which was computed for each location (ignoring age) and is assumed to impact the instantaneous reproductive number two weeks later. The lag was introduced to distinguish local saturation and information effects from immunity in the general population.

\subsection{Information on incidence}
Additionally, the historic case data based on reporting date (instead of symptom onset) allow to reconstruct the publicly available information about county specific case load. Reported cases are assumed to influence behaviour the following day. The logarithm of county-specific publicly known 7-day \emph{incidence} is used to represent this information about local risk of infection. Specifically the transformation $log_{10}(1+cases/pop \times 10^5)$ was used, where cases is the 7-day accumulated cases and pop the population of the location at hand. This variable obtains the value $0$ if no infections were reported in a week.

\subsection{Ratio of traced infectious}
Previous modelling studies suggested that delays between symptom onset and confirmation are important factors in outbreak control \cite{hellewell2020feasibility}. The data allows to construct for each day the ratio of local cases that has been reported to the health department among the potentially currently infectious cases. I refer to this ratio as the \emph{ratio of traced infectious}.

Individuals are assumed to be infectious one day prior to 6 days after symptom onset to avoid weekday effects. This assumption is in line with virus shredding \cite{he2020temporal} and contact tracing \cite{cheng2020contact} studies. As the health department is responsible for contact tracing the \emph{ratio of traced infectious} may hold substantive information about the immediacy of contact tracing and the speed of testing. 

Importantly, the ratio is computed based on the reported cases. The ratio of traced infectious \emph{among all infections} arises after dividing by the reporting rate. This has implications for the interpretation of the effect estimate of the ratio of traced infectious. The effect estimate of testing and tracing on an individual primary case arises after multiplying with the inverse of the reporting rate as illustrated in Section \ref{sec:supp_testing}.

\begin{figure}
	\centering
	\includegraphics[width=\linewidth]{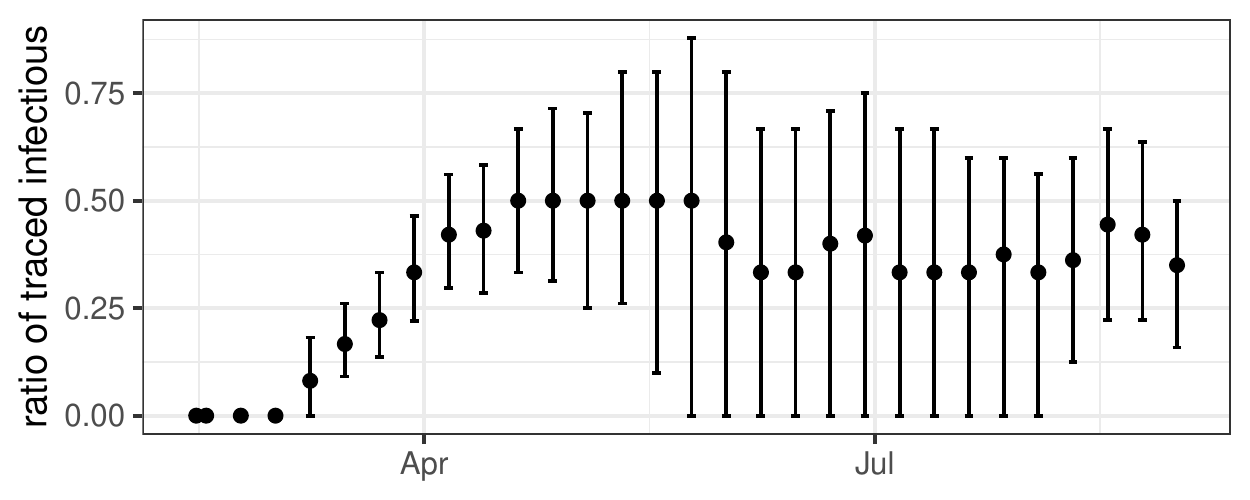}
	\caption{Median and $80\%$ confidence bands (across subregions) for ratio of infectious cases (one day before until 6 days after symptom onset) that are already reported and presumably subject to contact tracing. Data is aggregated by week.}
	\label{fig:traced}
\end{figure}

The development of the ratio of traced infectious over time is shown in Figure \ref{fig:traced}. In March, the average ratio increased from 0 to 30\%. By the end of April the ratio reached it's peak at 50\% before staying mostly constant for the remaining time. Importantly, the regional variation illustrated by the 80\%-confidence intervals is substantial.


\subsection{Weather}

Daily location specific weather data was obtained from the Climate Data Center (CDC) of the German Weather Service (Deutscher Wetter Dienst - DWD). For each county the temperature and relative humidity from any weather station within 50 kilometers was considered and daily averages were computed.

\begin{figure}
	\centering
	\includegraphics[width=\linewidth]{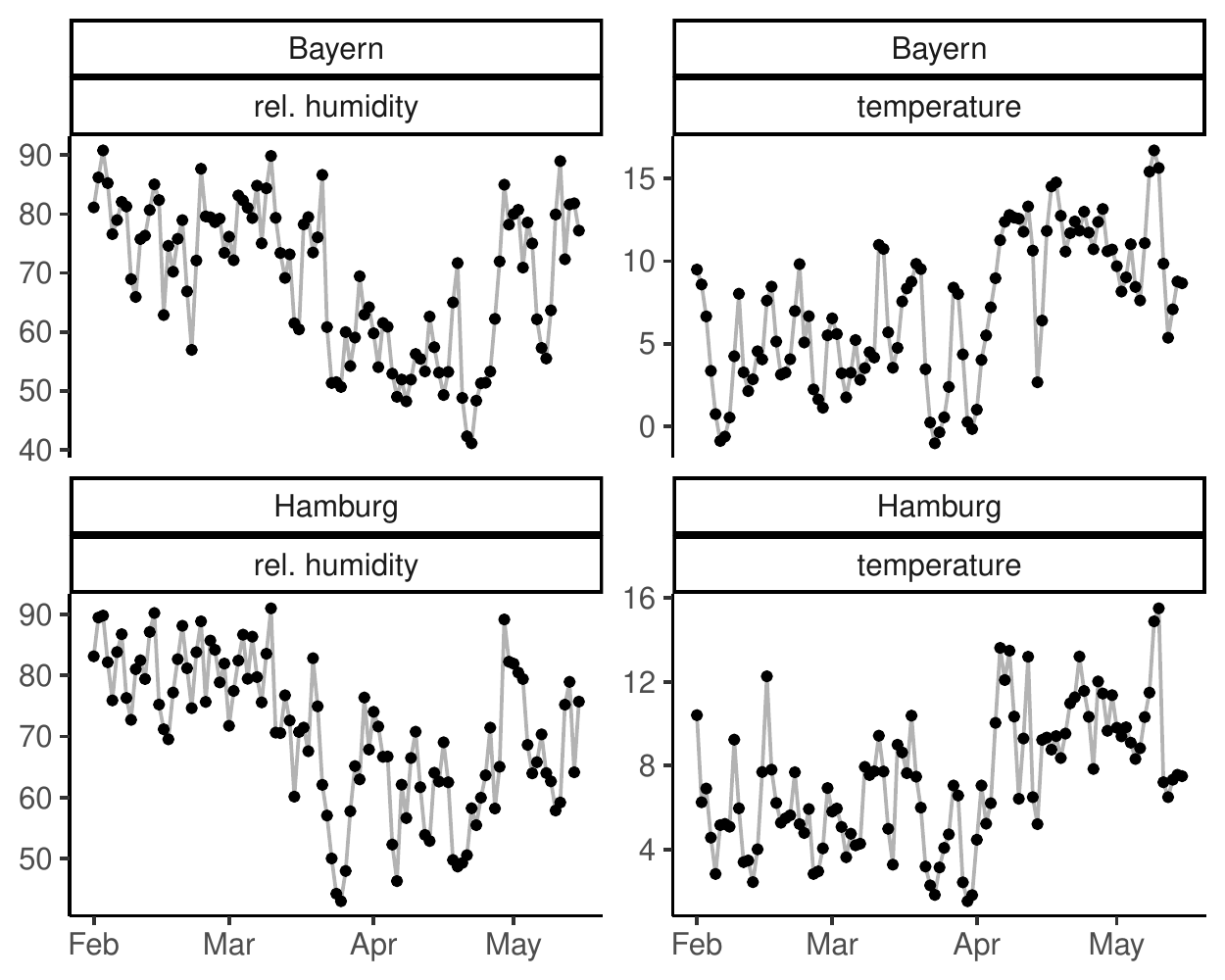}
	\caption{Daily time series of average temperature and relative humidity for two German states (Bayern and Hamburg). Temperature is given in degree Celsius. Relative humidity is given in percentage points.}
	\label{fig:weather_example}
\end{figure}

Figure \ref{fig:weather_example} plots example time series of weather covariates for two regions of Germany to illustrate the substantial variation in daily weather. Figure \ref{fig:cor_main} illustrates the covariance structure between covariates and shows that relative humidity and average temperature exhibit an empirical correlation of $-0.32$, which suggests that there is sufficient variation between the two weather variables to distinguish their associations with the instantaneous reproductive number.

A summary of all real-valued covariates can be found in Table \ref{tab:cov_real}. The variation of each covariate that is not associated with location or time can be considered most valuable for a robust identification of effects. Variation in location is captured by the basic reproductive number. Variation in time can be argued to be more susceptible to model misspecification and confounding unobservables. 

\begin{table}[ht]
	\centering
	\small
	\begin{tabular}{lrrrrrrr}
		\hline
		label & min & q05 & mean & q95 & max & R$^2$ (time) & R$^2$ (location) \\ 
		\hline
		cumulative incidence (\%) (r) & 0.00 & 0.00 & 0.10 & 0.39 & 1.26 & 0.47 & 0.26 \\ 
		incidence (logarithm) (r,s) & 0.00 & 0.00 & 0.96 & 2.01 & 2.73 & 0.82 & 0.06 \\ 
		average temperature (r,s) & -1.91 & 1.36 & 7.94 & 14.43 & 18.20 & 0.89 & 0.03 \\ 
		ratio of traced infectious (r) & 0.00 & 0.00 & 0.28 & 0.78 & 1.00 & 0.55 & 0.07 \\ 
		relative humidity (r,s) & 31.80 & 45.43 & 66.84 & 87.61 & 95.84 & 0.79 & 0.01 \\  
		\hline
	\end{tabular}
	\caption{Summary of real-valued covariates. Table denotes main covariates, their minimum, maximum, $5\%$- and $95\%$-quantile, mean, and the proportion of the variance (R$^2$) that is predictable by time and location respectively.}
	\label{tab:cov_real}
\end{table}

\subsection{Interventions}

Policy interventions were specifically catalogued for this study. The full data set with references can be accessed online. A descriptive summary of the most important interventions can be found in Table \ref{tab:cov_dummies}. Many intervention effect estimates rely almost entirely on variation across time. A description of each intervention can be found in Table \ref{tab:descr_intervention}.

\begin{table}[ht]
	\centering
	\begin{tabular}{lllrrrr}
		\hline
		label & first & days & location & R$^2$ (time) & R$^2$ (location) \\ 
		\hline
		public awareness rising & 2020-03-13 &  64 & 111 & 1.00 & 0.00 \\ 
		schools or daycare closed & 2020-03-14 &  63 & 111 & 0.98 & 0.00 \\ 
		restaurants closed & 2020-03-14 &  63 & 111 & 0.95 & 0.00 \\ 
		sports limited & 2020-03-14 &  63 & 111 & 0.95 & 0.00 \\ 
		gatherings forbidden & 2020-03-17 &  60 & 111 & 0.79 & 0.08 \\ 
		non-essential shops closed & 2020-03-17 &  60 & 111 & 0.93 & 0.00 \\ 
		stay-at-home order & 2020-03-20 &  50 &  39 & 0.18 & 0.25 \\ 
		distancing in public & 2020-03-21 &  56 & 111 & 0.97 & 0.00 \\ 
		narrow testing & 2020-03-25 &  52 & 111 & 1.00 & 0.00 \\ 
		masks in public & 2020-04-06 &  40 & 111 & 0.96 & 0.00 \\ 
		daycares reopen & 2020-04-20 &  26 & 111 & 0.64 & 0.06 \\ 
		schools reopen & 2020-04-20 &  26 & 104 & 0.59 & 0.07 \\ 
		churches reopen & 2020-04-24 &  22 & 111 & 0.85 & 0.01 \\ \hline
	\end{tabular}
	\caption{Summary of interventions. Table denotes main interventions, first implementation, total number of days intervention was active at any location, total number of locations that implemented intervention at some point, and the proportion of the variance (R$^2$) that is predictable by time and location respectively.}
	\label{tab:cov_dummies}
\end{table}

\begin{table}[ht]
	\centering
	\small
	\begin{tabular}{L{5cm}L{11cm}}
		\hline
		label & description \\ 
		\hline
		schools or daycares closed &  Either daycares or schools are closed. Partly open for key workers or vulnerable groups is also denoted as closed. Closing of individual schools if not enforced by regulations is ignored. \\ \hline
		Public awareness rising & First major public speeches of German President and health minster encouraging changes in behaviour. \\ \hline
		restaurants closed & Closing of restaurants. Limited capacity is not denoted as closed. Take-away only is denoted as closed. \\ \hline 
		gatherings forbidden & Any limitations on gatherings in public space. Changes in number of people are ignored. Common categories were one household only, two households only, or a limitation in group size to 5 or 10.\\ \hline
		non-essential shops closed & Closing of non-essential shops. Deviations in details (e.g., hardware stores) are ignored and all restrictions are denoted as closed. \\ \hline
		stay-at-home order & Regulations that mandated to stay at home unless exceptions arise. Exceptions always included work and individual sport.\\ \hline
		distancing in public & Mandated minimum distance in public space between individuals. Exceptions include often the own household or family. Common distance was 1.5 meters.\\ \hline
		narrow testing & The testing guidelines changed for the public and testing was limited to symptomatic cases that additionally had either exposure to another case or were travelling in high risk areas as denoted by the RKI. \\ \hline
		masks in public & Community masks obligatory for supermarkets and public transport. \\ \hline
		daycares reopen & Daycares reopen again. Often under limited capacity and with new safety concept. \\ \hline
		schools reopen & Schools reopen for all classes. Often under limited capacity and with new safety concept. If only particular classes were allowed, school is denoted as closed.\\ \hline
		churches reopen & Religious public gatherings were allowed. Often under limited group size and safety concept.\\ \hline
		symptomatic testing & The testing guidelines changed for the public and testing was available for any symptomatic person.\\ \hline
		\hline
	\end{tabular}
	\caption{Description of main interventions ordered by date of implementation. Full list of interventions and their description can be found in the accompanying data.}
	\label{tab:descr_intervention}
\end{table}

The responsibility for public health interventions lies mostly on the state level (Bundesland). Some policy measures (like testing regime) were decided on a national level. Many were implemented simultaneously after state leaders coordinated their response. A few counties (sub-regional level) deviated by imposing additional restrictions (e.g. earlier mandate for masks in public).

The data collection effort focused on the most impacted states Berlin, Bayern, Niedersachsen, Brandenburg, Nordrhein-Westfalen, Baden-Württemberg, Thüringen, Hessen, Hamburg, and Mecklenburg-Vorpommern. Additional interventions were found on the county level for LK Rottweil, LK Tirschenreuth, SK Leverkusen, LK Heinsberg, LK Coesfeld, SK Jena, SK Wolfsburg, and SK Braunschweig.

\begin{figure}
	\centering
	\includegraphics[width=\linewidth]{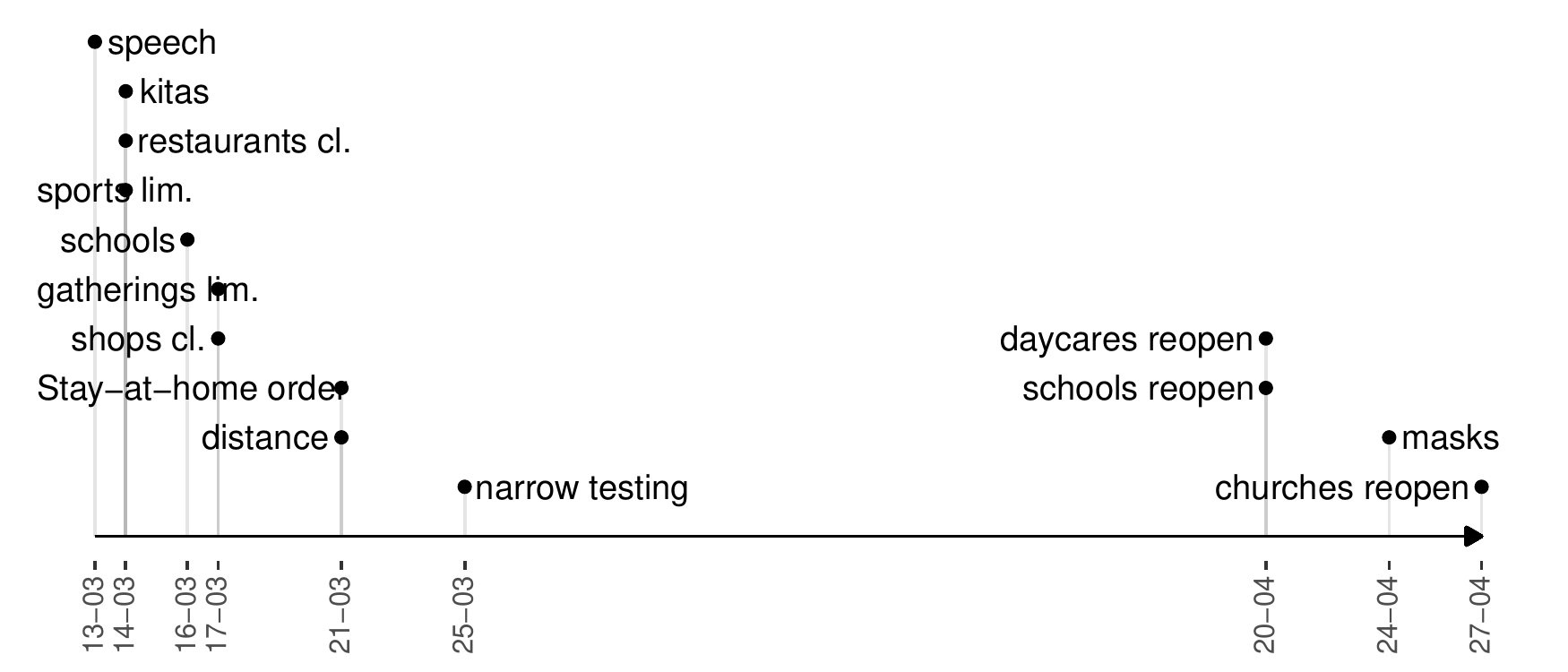}
	\caption{Timeline of important interventions. The first implementation on state level is shown, ignoring implementation on county level.}
	\label{fig:interventions_timeline}
\end{figure}

\begin{figure}
	\centering
	\includegraphics[width=\linewidth]{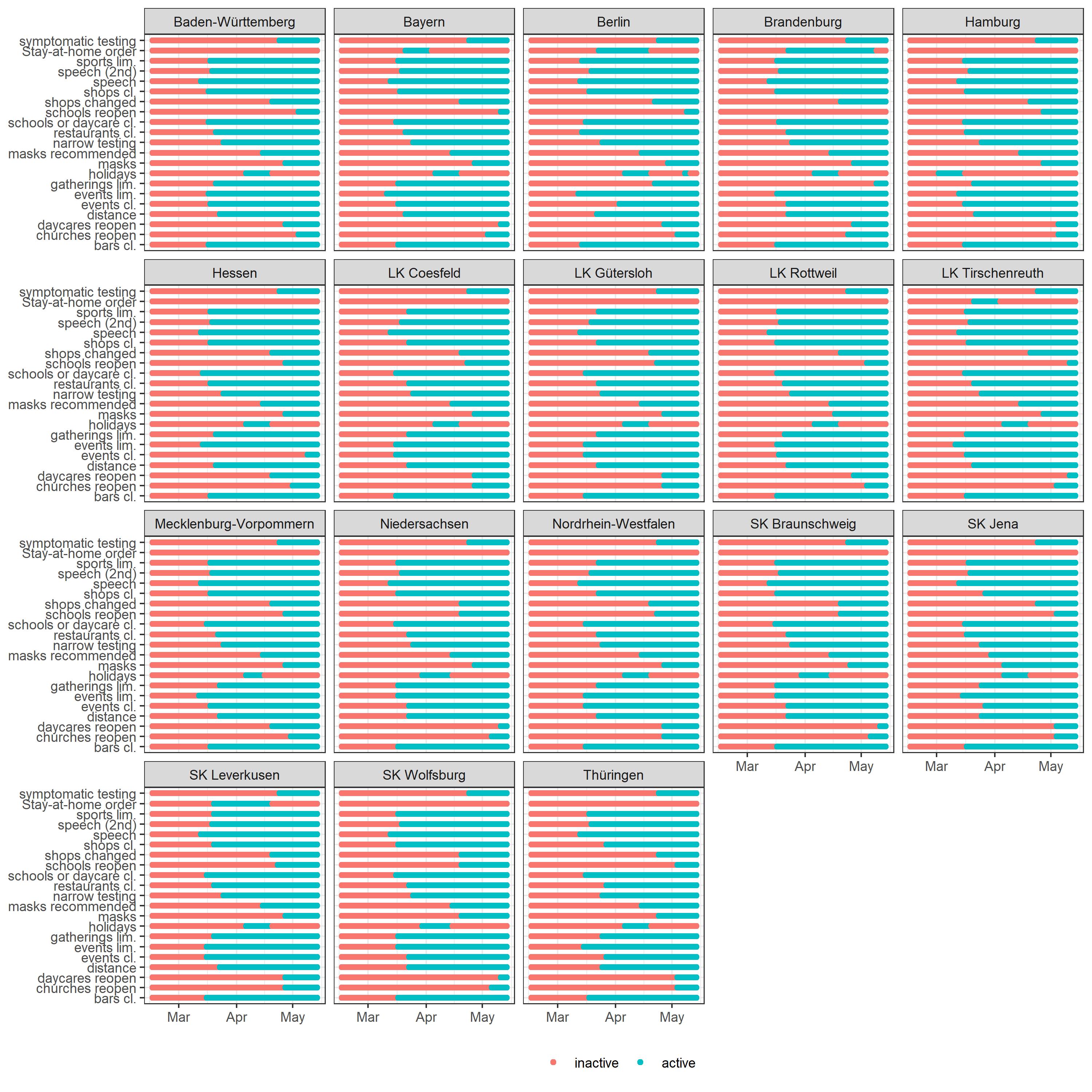}
	\caption{Intervention dummies by location. Interventions are shown for all states in the sample and all counties deviating from state interventions.}
	\label{fig:interventions_time_unit}
\end{figure}

\begin{figure}
	\centering
	\includegraphics[width=\linewidth]{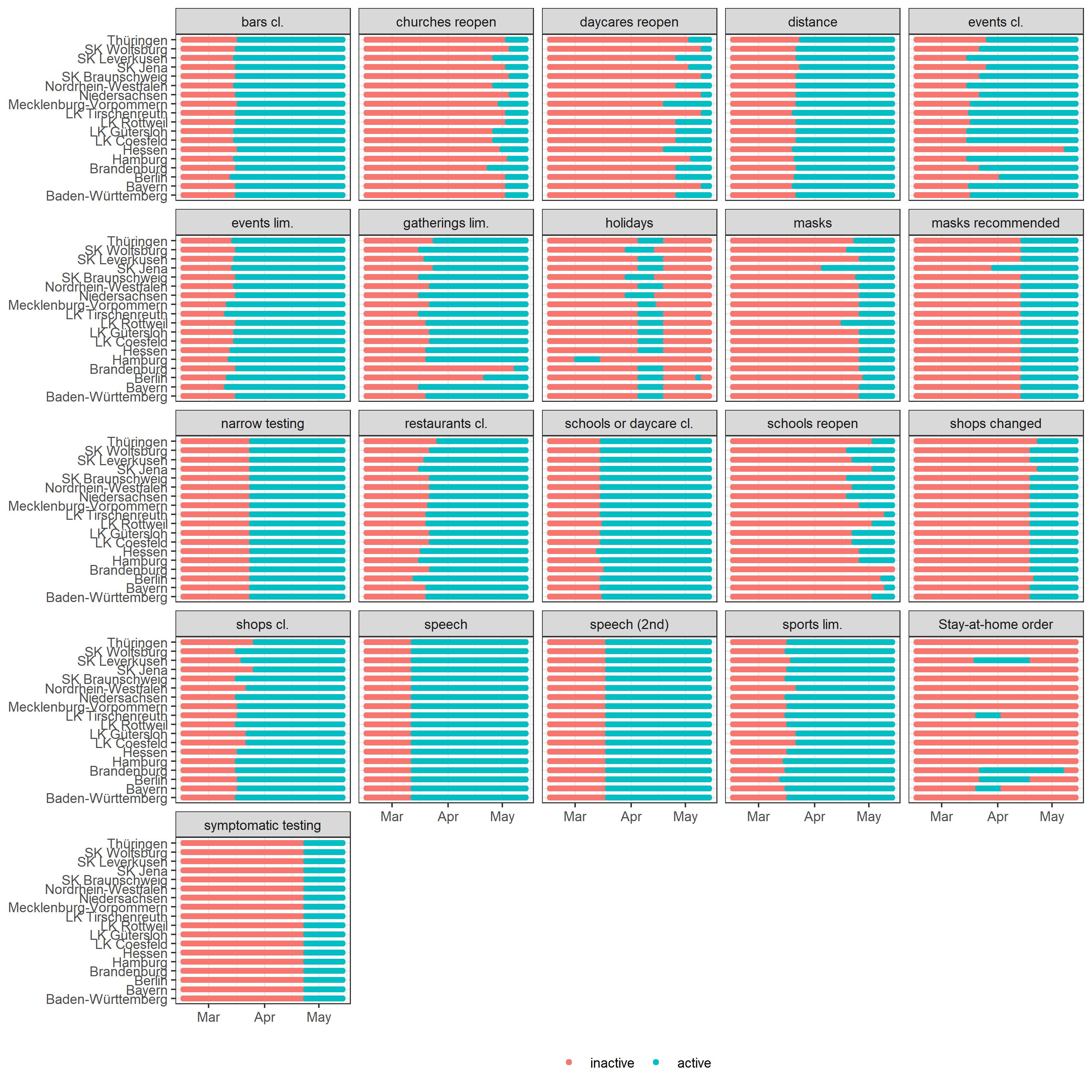}
	\caption{Intervention dummies by intervention. Interventions are shown for all states in the sample and all counties deviating from state interventions.}
	\label{fig:interventions_time}
\end{figure}

Information on the timing of the most important interventions is given as timeline in Figure \ref{fig:interventions_timeline} and further details can be found in Table \ref{tab:cov_dummies}. Full information on the timing of enactment in different locations is provided in Figures \ref{fig:interventions_time} and \ref{fig:interventions_time_unit}.

All interventions are coded as active, when they can be assumed to have impact. For example, the speeches given at March 12 are denoted as active starting March 13. If the closing of schools was announced \emph{entering into force immediately}, they were denoted as active the day after.

Closing of schools, daycare, shops, sports is still denoted as active when the respective reopening takes place. Thus, estimates of the reopening can be directly understood as the effect of reopening. One exception is the \emph{stay-at-home order}, which was lifted after a relatively short amount of time and not implemented in all states.

\section{Prior choice}\label{sec:supp_priors}

Priors were chosen with the goal to enable identification trough information in the literature, while allowing for adaptation in the context of the study. For effects, dispersion, and initial conditions weakly informative priors were chosen. Table \ref{tab:priors} lists all prior choices.

Generation time distribution $D_i$ and incubation period distribution $D_s$ are assumed to have a gamma shape with standard a deviation of $2$ and a mean that has a normal prior with a mean of $5.5$ days and a standard deviation of $0.1$. For a review on the incubation period see \cite{McAloone039652}.

Infections are initialized when the first symptomatic case was reported. The six previous days have a prior for initial infections that is exponentially distributed with mean $\frac{\mu_{init}}{6}$, where $\mu_{init}$ has a positive normal prior with mean $4$ (mirroring the reporting rate of $0.25$) and standard deviation $4$.

Effects of interventions and covariates are equipped with a normal prior with mean $0$ and standard deviation $0.2$. Additionally, the model has an error term with standard deviation $0.1$ to prevent that misspecification of $R_t^{l,a}$ is attributed to individual dispersion. Results without this error term are largely robust (not shown here).

\begin{table}
	\begin{tabular}{clll}
		parameter & description  & prior or parameter choice \\
		\hline
		$\beta^a_j$ & effect of covariate $j$ on age group $a$ & $\mathcal{N}(0,0.2)$\\
		$\beta^a_{t,l}$ & multiplicative error term for $R_t^a$ at location $l$ & $\mathcal{N}(0,0.1)$\\
		$r_t$ & reporting rate & $0.25$\\
		$D_i$ & generation time distribution  & Gamma shaped $\Gamma(\mu_i,\sigma_i)$ \\
		$D_s$ & incubation period distribution & Gamma shaped $\Gamma(\mu_s,\sigma_s)$\\
		$\mu_i$ & mean generation time & $\mathcal{N}(5.5,0.1)$ \\
		$\mu_s$ & mean incubation period & $\mathcal{N}(5.5,0.1)$ \\
		$\sigma_i$ & standard deviation generation time & 2 \\
		$\sigma_s$ & standard deviation incubation period & 2 \\
		$\Psi^a$ & dispersion parameter for age group $a$ & $\mathcal{N}^+(0,5)$\\
		$\mu_{init}$ & expected initial infections & $\mathcal{N}^+(4,4)$\\
		$d_{init}$ & number of days for initial infections & 6 \\
		$i_t$ & initial infections for $t=t_0,\dots,t_0+d_{init}$ & $exponential$ with mean $\frac{\mu_{init}}{d_{init}}$\\
	\end{tabular}
	\caption{Parameter and prior choices. $\mathcal{N}(\mu,\sigma)$ denotes a Gaussian distribution with mean $\mu$ and standard deviation $\sigma$. $\mathcal{N}^+(\mu,\sigma)$ the respective half-normal distribution.}
	\label{tab:priors}
\end{table}

\section{Results}

\subsection{Implementation of MCMC}
The data was prepared and the results were analyzed with the statistical software R 3.6.3 \cite{rmanual}. The MCMC sampler was constructed with JAGS 4.3 \cite{jagsmanual}. The burn-in phase was $10.000$ iterations. And $10.000$ iterations were sampled subsequently, which were then thinned to $1000$ draws for inference. The maximum Rhat-value among the monitored variables (excluding the latent infection process for computational reasons) was $1.16$ and visual diagnostics were administered to assess convergence. 

Replication code is available online.\footnote{\url{https://github.com/Schmidtpk/InfSup}}

\subsection{Transmission}\label{supp:res_transmission}

\begin{figure}
	\centering
	\includegraphics[width=.6\linewidth]{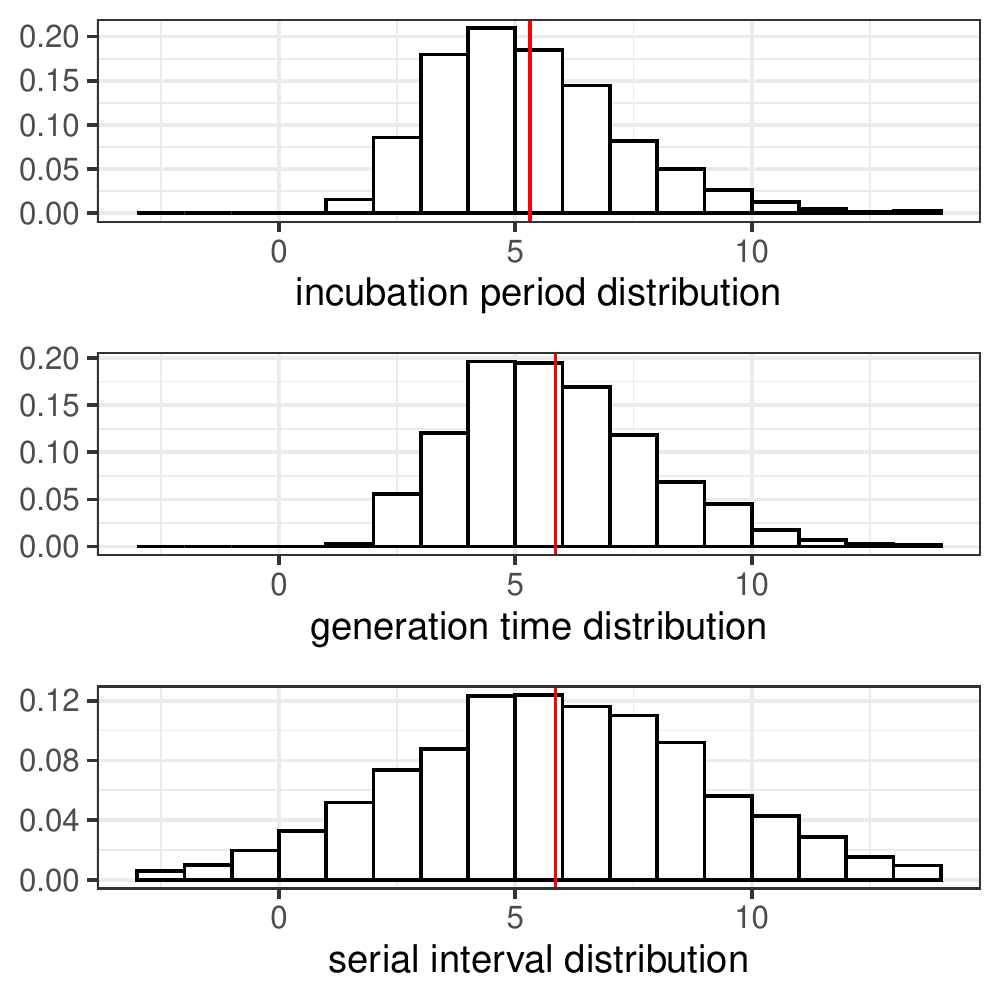}
	\caption{Incubation period distribution, generation time distribution, and serial interval distribution. The average is shown with a red line. Incubation period and generation time are estimated in the model. The serial interval is deduced by assuming independent generation time and incubation period.}
	\label{fig:dist}
\end{figure}

Results for incubation period and generation time are illustrated in Figure \ref{fig:dist}.

Noteworthy, symptom onset is self-reported and therefore prone to biases. In particular, patients are incentivized to report symptoms to increase their likelihood of receiving a test and full health insurance coverage for the associated expenses. Further, it can be assumed that symptom onset is often elicited after the positive test and unrelated symptoms are assigned as first symptoms, which would lead to an overestimation of the incubation period.

The basic reproductive number is estimated separately for each location and age compartment. Table \ref{tab:meta} shows a meta regression that explains variation in basic reproductive number, timing of first symptomatic case, and number of infections before the first case became symptomatic. 

The age averaged basic reproductive number (in the absence of interventions and under average macro conditions) is $2.53$ and positively associated with population density. In general, variation due to explanatory variables is very little ($R$-squared of $0.08$). As expected population density has a small positive effect. The ratio of 15-34 year olds has a small negative effect, which might indicate the reduced susceptibility of younger age groups. Variation in timing of initial (detected) exposure varies more prominently, where early initial infections are explained by high population density and a large ratio of young inhabitants. The accumulated effect of one standard deviation in population density and age groups amounts to about 6-7 days earlier initial infection. Given the basic reproductive number in winter, this accumulates to about 3-4 times as much initial exposure. Finally, the number of infections already ongoing, when the first case develops symptoms is on average 4.6 (driven by the assumption that the reporting rate $r_t$ is $0.25$) and no significant covariates were detected.

No evidence for a significant difference for Eastern Germany can be found, which suggests that the stark difference in incidence between Eastern and Western Germany is mostly driven by initial exposure and potentially population density.

\begin{table}[htp]
	\centering 
	\caption{Meta regression of locations. A regression of the mean of the basic reproductive number $R_0$, the timing of the first symptomatic case (days after 2020-02-15), and the number of infected when first case developed symptoms. Explanatory variables are population density, dummies for rural county and Eastern Germany (including Berlin), and ratio of different age groups. Continuous covariates are standardized. Basic reproductive number and number of initial infections are weighted average over age-specific results.} 
	\label{tab:meta} 
	\begin{tabular}{@{\extracolsep{5pt}}lccc} 
		\\[-1.8ex]\hline 
		\hline \\[-1.8ex] 
		& \multicolumn{3}{c}{\textit{Dependent variable:}} \\ 
		\cline{2-4} 
		\\[-1.8ex] & $R_0$ & timing first case & number of infections \\ 
		\hline \\[-1.8ex] 
Eastern Germany & $-$0.042 & $-$0.224 & $-$0.492 \\ 
& (0.059) & (1.284) & (0.307) \\ 
& & & \\ 
population density & 0.072$^{**}$ & $-$1.764$^{**}$ & 0.235 \\ 
& (0.034) & (0.743) & (0.177) \\ 
& & & \\ 
rural (no city) & 0.075 & $-$2.815 & 0.182 \\ 
& (0.080) & (1.739) & (0.415) \\ 
& & & \\ 
ratio age group 15-34 & $-$0.057$^{*}$ & $-$2.808$^{***}$ & 0.033 \\ 
& (0.032) & (0.694) & (0.166) \\ 
& & & \\ 
ratio age group 35-59 & 0.005 & $-$2.437$^{***}$ & $-$0.077 \\ 
& (0.032) & (0.705) & (0.168) \\ 
& & & \\ 
average & 2.531$^{***}$ & 28.989$^{***}$ & 4.611$^{***}$ \\ 
& (0.061) & (1.334) & (0.318) \\ 
& & & \\ 
\hline \\[-1.8ex] 
Observations & 111 & 111 & 111 \\ 
R$^{2}$ & 0.068 & 0.277 & 0.087 \\ 
Residual Std. Error (df = 105) & 0.214 & 4.675 & 1.116 \\  \hline \\[-1.8ex] 
		\textit{Note:}  & \multicolumn{3}{r}{$^{*}$p$<$0.1; $^{**}$p$<$0.05; $^{***}$p$<$0.01} \\ 
	\end{tabular} 
\end{table}

\subsection{Policy interventions}

\begin{figure}
	\includegraphics[width=.9\linewidth]{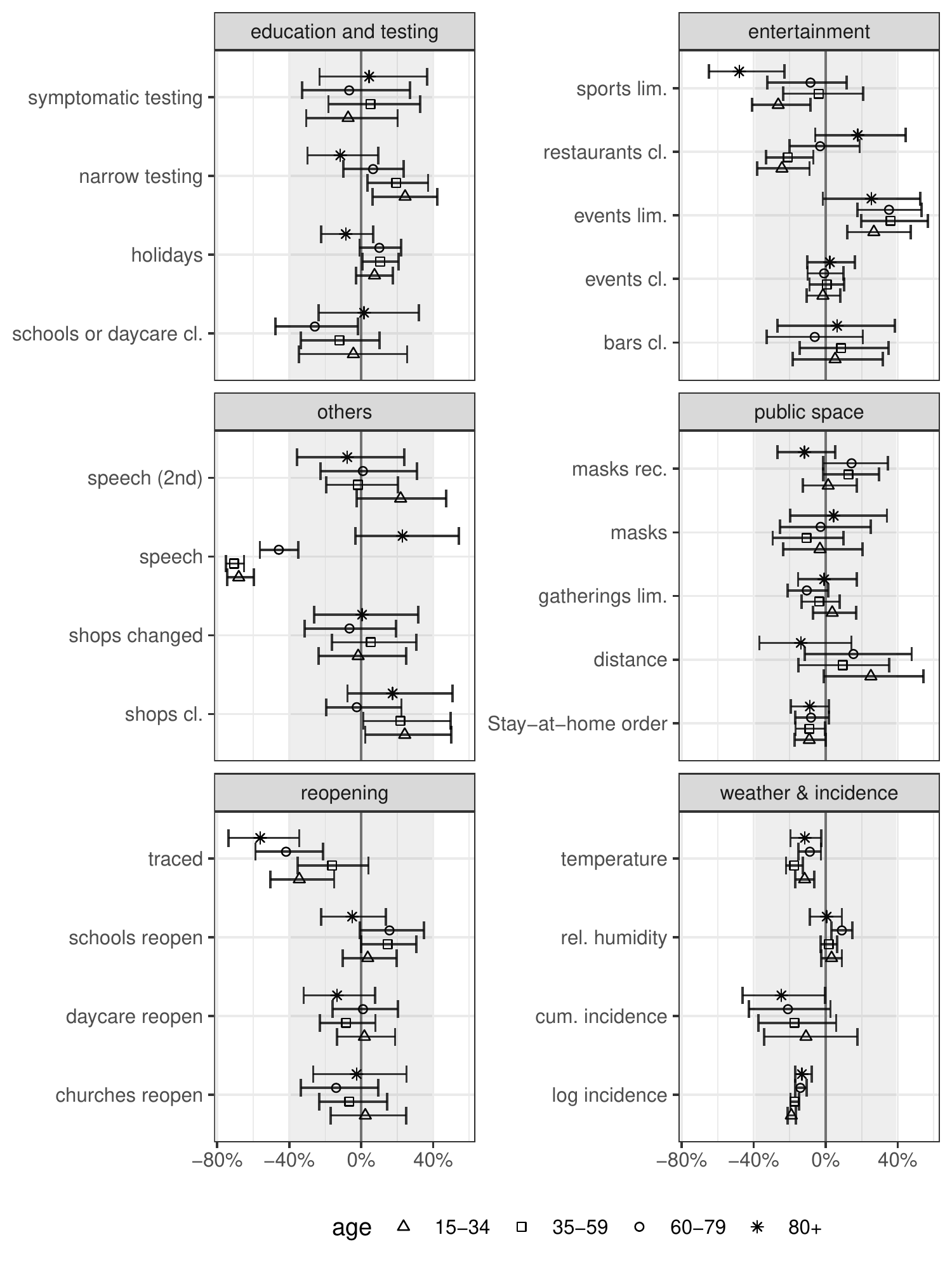}
	\caption{Changes in transmission. The plot depicts average effects and $95\%$ confidence intervals for different age groups for all covariates excluding fixed effects. The grey shade denotes the respective confidence interval of the prior. Descriptions use abbreviations for closed (cl.), limited (lim.), recommended (rec.), and cumulative incidence (cum. inc.).}
	\label{fig:effects}
\end{figure}

In the following the age specific effect estimates shown in Figure \ref{fig:effects} are discussed. For additional details on interventions see Table \ref{tab:descr_intervention}. All effect estimates should be understood as average changes in transmission associated with the situation the interventions were implemented in. It is subject to discussion, if the same effect can be expected to manifest itself under different circumstances.

Average effects of covariates are shown in the main text in Figure \ref{fig:effects_simple} and are based on the German age distribution. As age-specific effects for any single intervention were uncorrelated, the marginal effects are estimated more sharply than the age-specific effects. In the following, the most important differences in age are discussed.

Narrow testing, reducing the availability for tests to risk groups, health care workers and individuals exposed to a confirmed case, increased transmission to younger age groups, and reduced transmission to older age groups. The return to symptomatic testing end of April mitigates those differences. 

Holidays show some evidence for increased transmission in younger age groups, but decreased transmission in the age group above 80 years. 

The closing of schools and daycares was associated with a reduction in all age groups, except above 80 years. Noteworthy, there is little evidence in the age group 15-34. The lack of evidence for children below 15 years complicates drawing any decisive conclusions.

Limiting sport activities is associated with a decrease in transmission for the age group 15-34. Surprisingly, an even larger effect is found for the age group 80+, which might be due to misspecification of the model. Elderly care was subject to substantial changes during the pandemic and making those changes available as data set is beyond the scope of this study, but should be considered for future work.

Restaurant closure is associated with a moderate decrease in transmission, where older age groups show no evidence for changes. Bar closures were mostly applied in parallel. The effect estimate is mostly driven by the small implementation differences in mid-March, where bar closure arguably had very little potential to reduce bar attendance. Closing and limiting of events shows no evidence for reducing transmission. In fact, for the first recommendations the association is positive, which might stem from the fact that early transmission was mostly due to importation from high risk areas, which is subject to different dynamics. The other policy interventions were implemented in an environment were cases dynamics were mostly driven by local infections.

The opening of schools, daycares, and allowing religious gatherings is associated with little to no increase in transmission. A small increase in transmission associated with school openings can be detected for the age groups 35-79. Noteworthy, all openings were under safety concepts adapted to the risk of SARS-CoV-2 transmission.

The first speeches of the German president and health minister asking for major behavioural changes are associated with a strong reduction in transmission for all age groups except for the age group $80+$. The second major speech by the chancellor Angela Merkel shows no effect on average, but is associated with some increase in the younger population. It should be noted that other interventions that are not in the data, might factor into the reduction associated with the first speech. This includes a reduction of importation of cases from international travel as the RKI denoted a number of European areas as risk areas and quarantine rules for homecoming tourists with symptoms were set in place.

The closing of shops is associated with an increase in transmission for younger age groups. Related, mandatory distance in public spaces is associated with an increase in younger age groups, which suggests that the reduction of public interactions might have been substituted with private interaction that was subject to higher risk of transmission. The possibility should be considered that especially younger individuals, who face less individual risk, are prone to increasing transmission, when public interaction is substituted by private interaction. In line with this thought, a stay-at-home order (with exceptions including individual sport and work) reduces transmission for all age groups.

\subsection{Testing and Tracing}\label{sec:supp_testing}
The ratio of traced cases has a strong impact on all age groups, which is especially strong for older age groups. This suggests that a tracing reduces transmission to older more vulnerable groups. Naturally, this variable conflates two different effects that cannot be distinguished: Individual behavioural changes due to the positive test, and the effect of contact tracing as administered by the health department. 

Further, the ratio only measures the traced cases among reported cases, not among all infections. If the effect is assumed to be the same among the individuals not reported, the effect testing and tracing has at the individual level can be extrapolated by dividing the effect estimate by the reporting rate. This allows to infer the relative reduction in transmission for an infectious individual by testing and tracing.

\begin{figure}
	\centering
	\includegraphics[width=\linewidth]{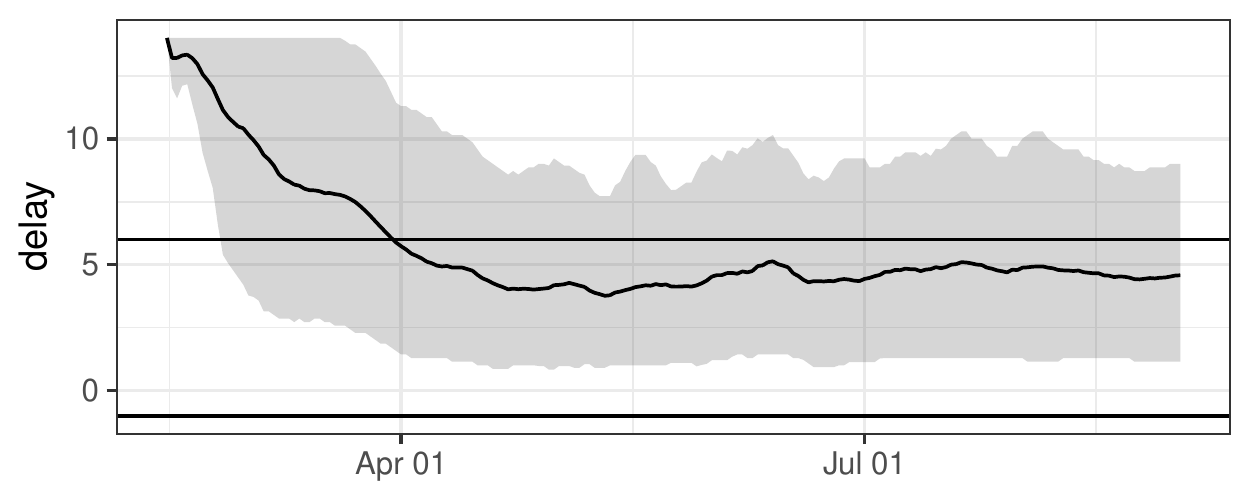}
	\caption{Average and $80\%$ confidence bands on delay between symptom onset and day of reporting. Delay is measured in days until first notice of local health department. The horizontal lines denote the critical phase of one day before until 6 days after symptom onset, which is hypothesized to be the days of highest infectiousness.}
	\label{fig:delay}
\end{figure}

It should be noted that most cases are reported between 2 to 6 days after symptom onset as shown in Figure \ref{fig:delay}. The effect estimate of tracing therefore is driven mostly by this region. It may be possible that earlier testing and tracing is un-proportionally more (or less) effective.

\subsection{Information and cumulative incidence}

Publicly reported local incidence decreased transmission. The effect is sharply estimated and stronger for younger age groups. 

Cumulative incidence is associated with a stronger reduction for old age groups. Cumulative incidence is given in percentage points. Thus, under random mixing, full immunity, and the absence of underreporting an effect of $1\%$ should be expected.

\subsection{Weather}

High average temperature reduced transmission. This effect is consistent across age groups. In the model period average temperature ranged from $-2^{\circ} C$ to $18^{\circ}C$ ($0.95$-CI: $[0^{\circ} C, 15^{\circ} C]$).

Relative humidity is significant on average and for the age group 60-79. The potential role of relative humidity in the transmission dynamics of SARS-CoV-2 is discussed in \cite{ahlawat2020anoverview}. The results here suggest a minor role of relative humidity in Germany. External validity for the effect of relative humidity beyond the range observed here ($0.95$-CI: $[43\%, 90\%]$) is in doubt.

\subsection{Robustness check with more recent data}

As a robustness check the model was applied to data from May to September 2020. The instantaneous reproductive number $R_t$ is modelled as a function of weather variables and of the ratio of traced infectious and local information as before. As there is not sufficient data on regional interventions at this point in time, those were substituted by week fixed effects. The model also contains weekday fixed effects and noise terms as before.

In this model, the age group 5-14 years was included, as schools were partly open and the ratio of younger age groups among the cases increased compared to the low incidence in the first surge of cases in March and April.

Similar to the main study, the mean incubation period was found to be slightly less than mean generation time with $4.6$ and $6.6$ days respectively. The age specific results can be found in Table \ref{tab:transmission_late}. While the age groups 15-34 and 35-59, still show a higher tendency for superspreading, overall the dispersion estimates are much lower, indicating that superspreading events have become less common after April.

\begin{table}[ht]
	\centering
	\begin{tabular}{lrrrrrrrr}
	& \multicolumn{2}{c}{\textit{$R_0$}} & \multicolumn{2}{c}{\textit{dispersion $\Psi$}} & \multicolumn{2}{c}{ratio infecting} & \multicolumn{2}{c}{ratio from $20\%$}\\ 
\cline{2-3} \cline{4-5} \cline{6-7} \cline{8-9}
age & mean & sd & mean & sd  & mean & sd & mean & sd \\ 
\hline
		5-14 & 1.09 & 0.32 & 12.09 & 2.97 & 0.62 & 0.03 & 0.52 & 0.01 \\ 
		15-34 & 1.24 & 0.17 & 6.01 & 2.36 & 0.60 & 0.04 & 0.54 & 0.02 \\ 
		35-59 & 1.25 & 0.17 & 6.03 & 2.44 & 0.60 & 0.04 & 0.54 & 0.02 \\ 
		60-79 & 1.14 & 0.26 & 13.72 & 2.90 & 0.62 & 0.04 & 0.52 & 0.01 \\ 
		80+ & 1.04 & 0.35 & 9.31 & 2.85 & 0.61 & 0.03 & 0.53 & 0.01 \\ 
		\hline
	\end{tabular}
\caption{Estimated basic reproductive number $R_0$, dispersion $\Psi$, respective ratio of primary infections actually infecting secondary infections, and ratio of secondary infections from $20\%$ most infecting primary cases for different age groups. Results based on data from May to August 2020. The ratios of secondary infections was computed assuming a constant reproductive number of $1$.}
\label{tab:transmission_late}
\end{table}

\begin{figure}
	\includegraphics[width=\linewidth]{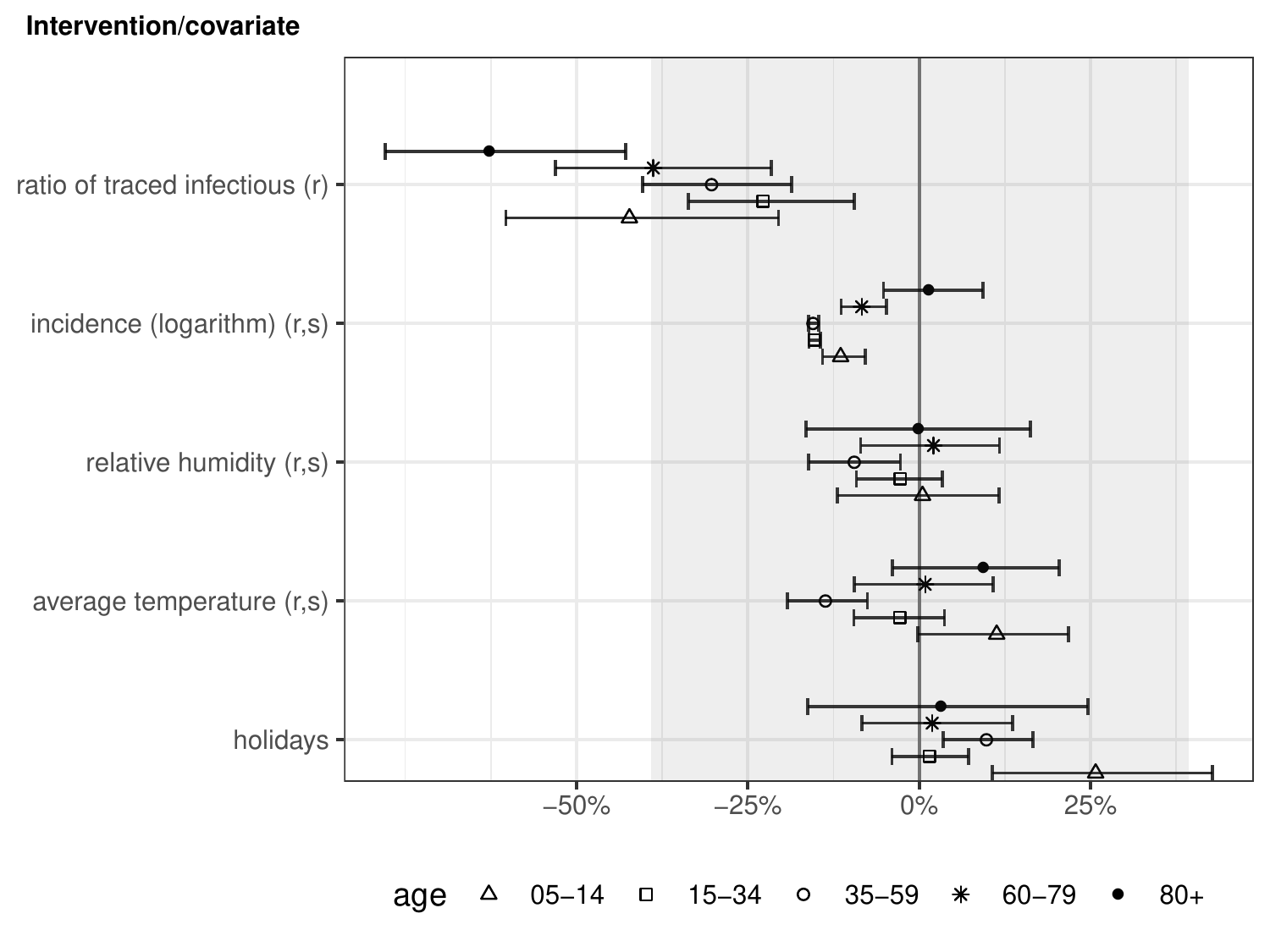}
	\caption{Effects on reproductive number for model applied to data from May to August 2020. The plot depicts average effects and $95\%$ confidence intervals for different age groups for all covariates excluding fixed effects. The shaded are indicates the $95\%$ confidence intervals of the prior distribution.}
	\label{fig:effects_late}
\end{figure}

The effect estimates for the covariates can be seen in Figure \ref{fig:effects_late}. The results are largely consistent with the main study. Testing and tracing is associated with a strong reduction in transmission. The effect is strongest for old and the youngest age group. This effect is slightly stronger, which might indicate that the reporting rate is higher as this would make the ratio of traced cases a better proxy for the ratio of traced infections.

Information on local incidence is again found to be a strongly correlated with a reduction in transmission. As before, this effect is mostly found among younger age groups. The age group 80+ shows no evidence for a successfully adaptation to current risk of infection.

\begin{figure}[htp]
	\centering
	\includegraphics[width=.8\linewidth]{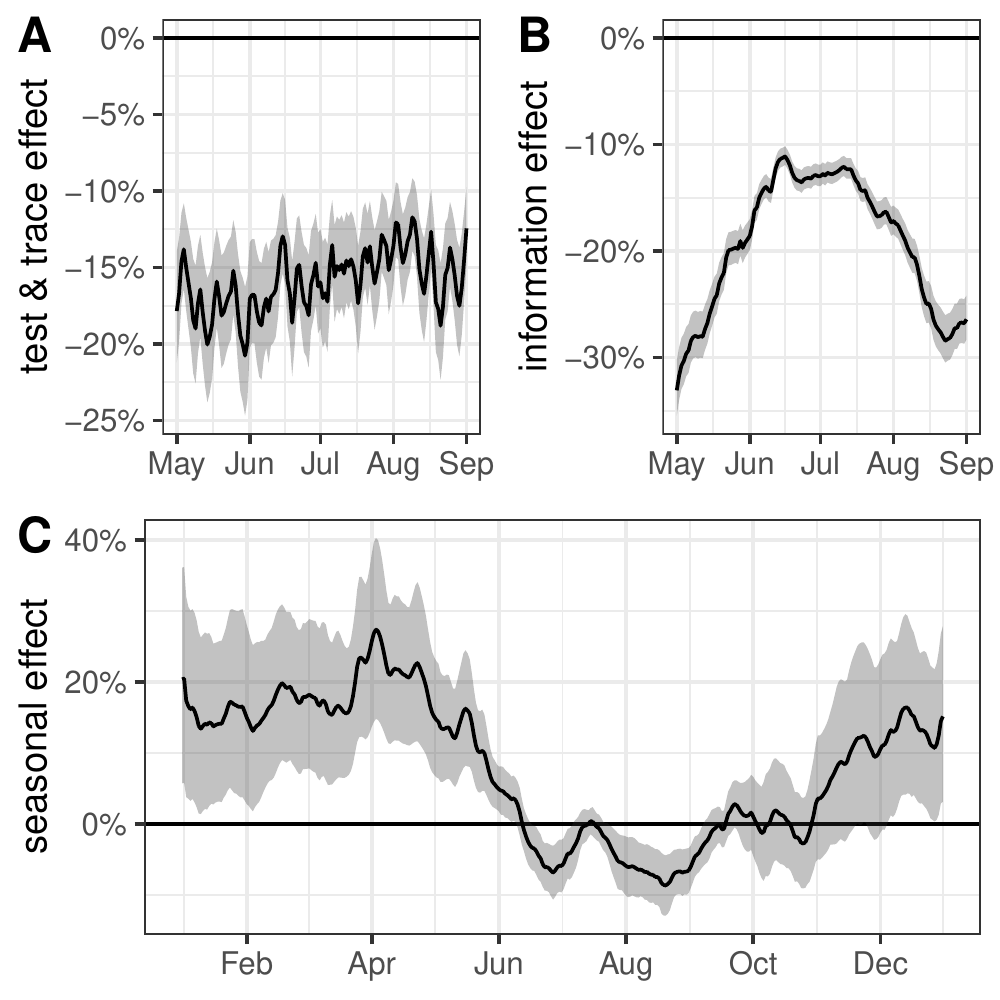}
	\caption{Total effect of testing and tracing (ratio of traced infectious), information (logarithm of reported local incidence), and season (average temperature and relative humidity). Estimates based on age group 15-59. $95\%$ confidence bands are shown. Figure A and B denote total effect given the data. Figure C \emph{extrapolates} the total effects of weather variables in an out-of-sample prediction based on average daily weather in the past three years, where confidence bands represent uncertainty in effect estimation, and results are smoothed with a 14-day rolling average.}
	\label{fig:total_all_late}
\end{figure}

Average temperature shows a similar pattern for the age groups 15-34 and 35-59. For the other age groups no significant effect of weather can be detected. Arguably, transmission to older and younger age groups is mostly driven by the middle age groups in professional or household setting, which would explain the pattern. The predicted seasonal effect of weather (Figure \ref{fig:total_all_late}) is smaller than in the main study, which suggests non-linear effects of temperature and relative humidity.

Holidays have a high variation state by state in Germany. Interestingly, there is a strong positive effect on school children below the age of 15, which suggests that school is a less risky environment for transmission than holidays. It should be noted that free tests were available at the border for German citizens when returning from international vacation, which might lead to a higher reporting rate for school children during holidays and an upward bias in the effect of holidays.

\section{Assumptions}\label{supp:assumptions}

In the following, the main assumptions of the model are listed.

\subsection{Reporting rate}

The key assumption to identify the reproductive number from report data is a correctly modelled probability of reporting $r_t^{l,a}$. In the main specification the reporting rate $r_t^{l,a}$ is assumed to be constant over time.

\begin{figure}[htp]
	\centering
	\includegraphics[width=\linewidth]{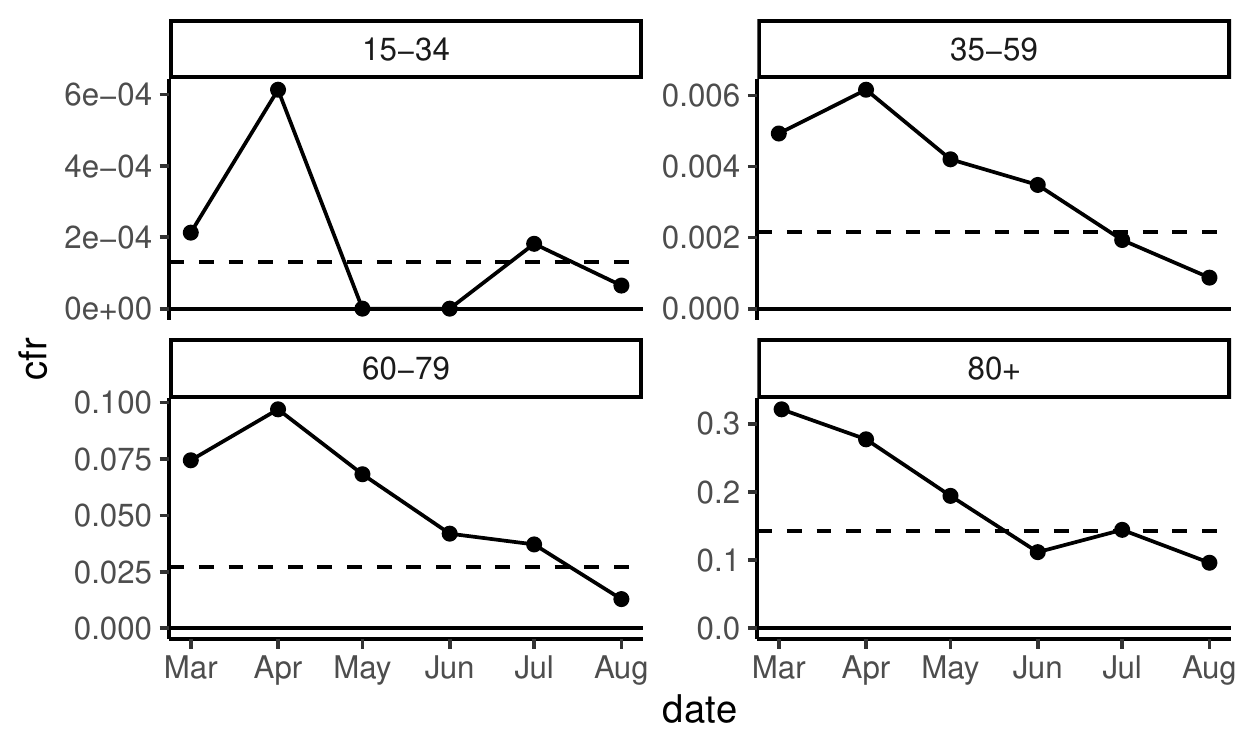}
	\caption{Case fatality rate over time for different age groups. 
		Ratio of deceased among cases first reported in each month.
		Vertical dashed line denotes age specific expected infectious fatality rate based on \cite{levin2020assessing}.}
	\label{fig:cfr}
\end{figure}

Previous studies argued that deaths are more reliable than case data \cite{flaxman2020estimating}. One limitation of death data is that identification of growth rates relies on a constant fatality rate, which is violated if age groups are not taken into account. As shown in Table \ref{tab:ftr} depicting German data, $27\%$ of cases over 80 years died, while only $0.03\%$ of cases under 35 years. If age groups are affected differently by interventions, relying on death data could induce strong biases. Effects on younger age groups are essentially undetectable. Further, improved hospital care is likely to have reduced the infected fatality rate over time. Figure \ref{fig:cfr} shows the symptomatic case fatality rate. The standard case fatality rate includes asymptomatic cases and requires adjustment by time as cases are often reported earlier than deaths. The RKI case data allows to compute fatality rate by symptom onset. The symptomatic case fatality rate is less prone to changes in testing and suggests that the infection fatality rate decreased over time.

\begin{table}[htp]
	\centering
	\begin{tabular}{lrrr}
		\hline
		age & cfr (in $\%$) & deaths & cases \\ 
		\hline
		A00-A04 & 0.03 &   1 & 3094 \\ 
		A05-A14 & 0.00 &   0 & 6883 \\ 
		A15-A34 & 0.03 &  19 & 55767 \\ 
		A35-A59 & 0.50 & 429 & 85859 \\ 
		A60-A79 & 8.16 & 2979 & 36485 \\ 
		A80+ & 26.96 & 5805 & 21534 \\ 
		\hline
	\end{tabular}
	\caption{Case fatality rate (cfr), cases, and deaths by age in Germany.}
	\label{tab:ftr}
\end{table}

Naturally, using case data is subject to a similar critique, as detection rates of infections may change over time. One advantage of using symptom onset instead of reporting date to aggregate case data, is that changing detection rates of asymptomatic cases does not impact the results.

The number of executed tests in Germany is published weekly by the RKI \cite{Seifried2020Erfassung}. Consistent reporting of those numbers started mid March. Between mid-March and mid-June there was only moderate variation in weekly tests between $327$ and $431$ thousand tests each week. Early test data is not available and case data may suffer from significant changes in the testing regime, which in turn may impact effect estimates for early interventions. In summer 2020 the test numbers increased further and reached a million each weak in August.

\begin{figure}
	\centering
	\includegraphics[width=.8\linewidth]{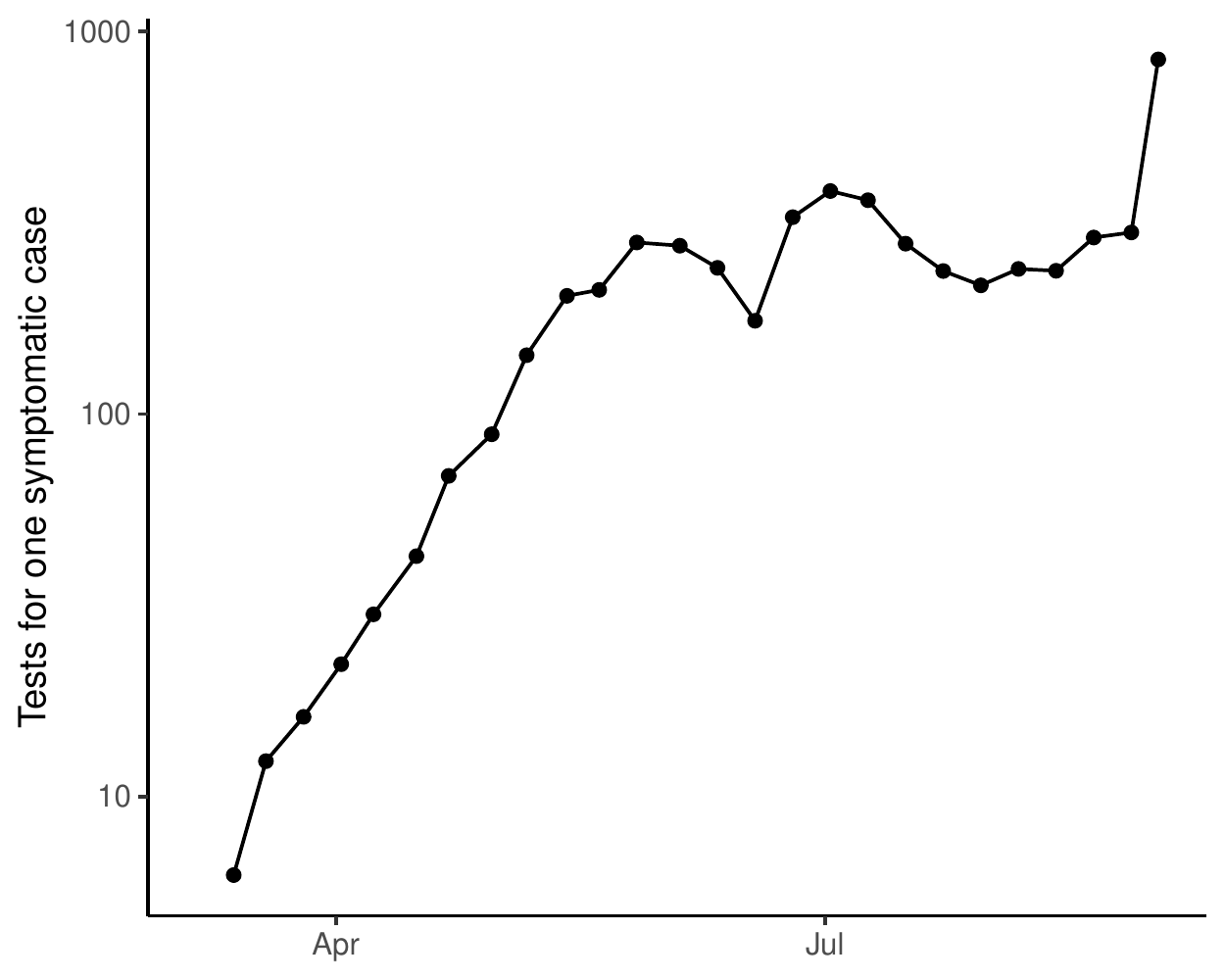}
	\caption{Average tests in a week for one reported case with symptom onset in week. Weekly number of tests as reported by the RKI and obtained from \cite{owidcoronavirus}.}
	\label{fig:tests}
\end{figure}

Figure \ref{fig:tests} shows the number of tests necessary to find one symptomatic case. Noteworthy, this number increased in spring (where test numbers stayed largely constant) and stayed constant in summer (where test numbers increased).

It was hypothesized that interventions reducing the doses of viral inoculum, e.g. face masks, increase the likelihood of asymptomatic caes \cite{gandhi2020facial}. Focusing on symptomatic cases, captures this effect. If asymptomatic cases are equally (or more) likely to spread, this could lead to a higher perlocution rate in the second generation that would not be captured by the model.

\begin{figure}
	\centering
	\includegraphics[width=.8\linewidth]{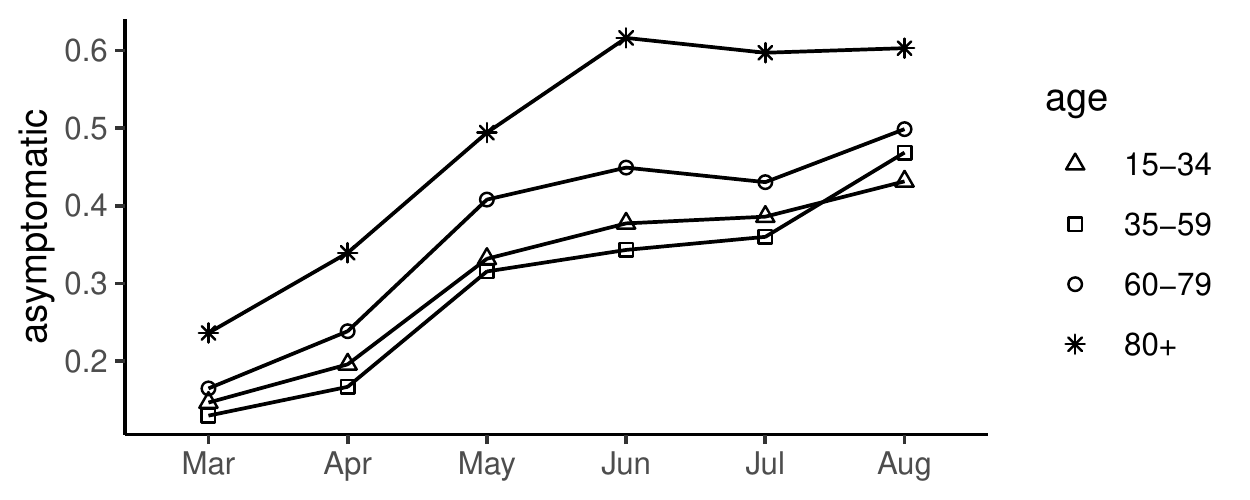}
	\caption{Ratio of asymptomatic cases over time.}
	\label{fig:asymptomatic_time}
\end{figure}

As can be seen in Figure \ref{fig:asymptomatic_time}, the ratio of reported cases without symptom onset increased over time. If the likelihood of developing (and reporting) symptoms remains constant over time, this suggests that more asymptomatic cases were found over time. In this situatoin, relying on reported cases instead of symptomatic cases might bias inference on growth rates.

\subsection{Absence of importation}\label{supp:importation}

The model proposed here ignores importation. Naturally, transmission between locations and age groups occur. Arguably, the obtained instantaneous reproductive number $R^{l,a}_t$ should only be read as reduced form summary of the current growth rate. Extension of the model incorporating importation across compartments are straight forward but require strong additional assumptions for the identification of reproductive numbers. Consider the transmission model
$$i_t^{l,a} \sim NB(\sum_{a'} L_t^{l,a'} R_t^{l,a,a'}, \sum_{a'} L_t^{l,a'} \Psi^{a}),$$
where $R_t^{l,a,a'}$ denotes the reproductive number from age group $a'$ to age group $a$. If detection rates are heterogeneous, identification of $R_t^{l,a,a'}$ requires that the ratio of detection rates between age groups is known. The implementation of such approaches would require reliable prevalence data that allow to identify detection rates of the German reporting system for PCR-positive test stratified by age and location.

\begin{figure}
	\includegraphics[width=\linewidth]{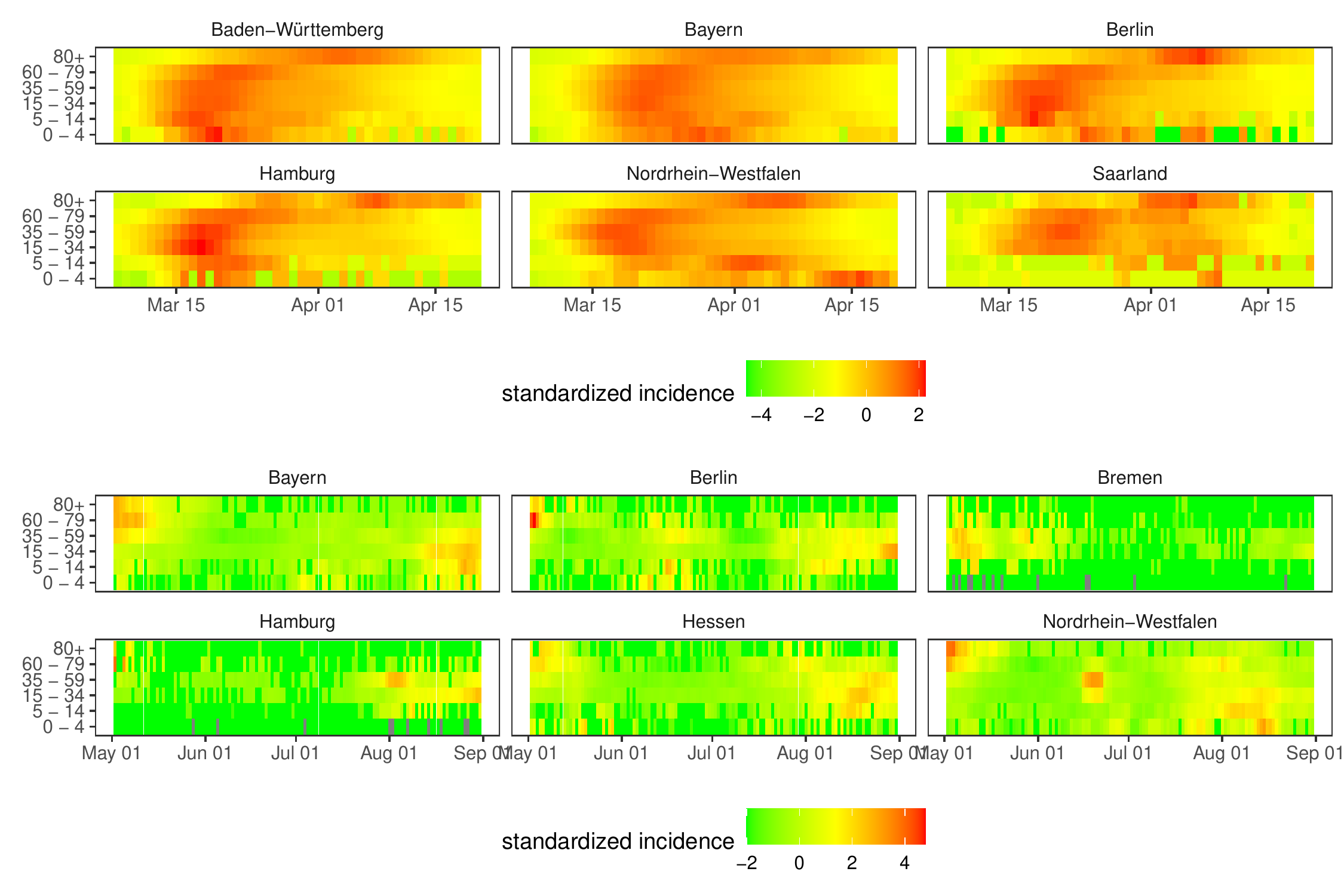}
	\caption{Standardized incidence heatmap for age groups over time. Incidence is standardized within age and location compartment.}
	\label{fig:age_heats}
\end{figure}

The data provides mixed evidence for transmission dynamics between age groups. Figure \ref{fig:age_heats} shows 7-day incidence based on symptom onset for the six regions with the highest incidence in Germany. The data is standardized to allow for different detection rates within age and location. While the changes in across age groups suggest that moderate age groups infected the elderly and children in the high incidence phase in March and April (distancing rules active and schools closed), the pattern becomes unclear in later stages during summer.

\subsection{Homogeneity}

Under the assumption of random mixing, i.e. each infected is equally likely to infect any other member of the population, the instantaneous reproductive number $R_t^{l,a}$ can be interpreted as a random draw from the current transmission situation in the population. In a more realistic setting, transmission is heterogeneous. A contact tracing study from Hong Kong finds, for example, that infections in social settings were associated with more secondary cases compared to household infections \cite{adam2020clustering}.

As transmissions occur more likely within a cluster (household, workplace, location, ethnicity, social class, etc.) $R_t^{l,a}$ can expected to exhibit auto-correlation in its error to depict an accurate description of the current \emph{average transmission dynamic} in the entire population. For the same reason, local immunity acquired within a cluster of infections (e.g., within a household) may temporarily underestimate the expected transmission dynamic in the entire population which will govern the process when infections are more equally distributed.

Such heterogeneity in transmission, as modelled for example via a spatial 
\cite{keeling1999effects} or a network structure 
\cite{davis2008abundance}, can lead to underestimating the effect of interventions that have a stronger effect on intra-cluster/long distance transmission. The reduced transmissions by the second generation of infections occur later and would not be attributed correctly to the initial intervention. 

Similarly, interventions might be more/less effective over time. Especially, the weather variables, which are identified based on daily variation in weather, are subject to the critique that long-term effects might differ substantially from short term effects.

Local saturation (e.g. in households or local communities) through immunisation can also play a role \cite{grassly2008mathematical}. The usage of a high number of subregional compartments can only partly control for that. 

\subsection{Unobservables}

Ultimately any transmission dynamics can be attributed to individual behaviour (potentially in interaction with external factors). The covariates considered here to explain the instantaneous reproductive number may omit other shared drivers of individual behaviour. One additional factor that may drive the spread of SARS-CoV-2 is the prevalence of new variants of the virus \cite{korber2020tracking}.

\begin{figure}[htp]
	\centering
	\includegraphics[width=.9\linewidth]{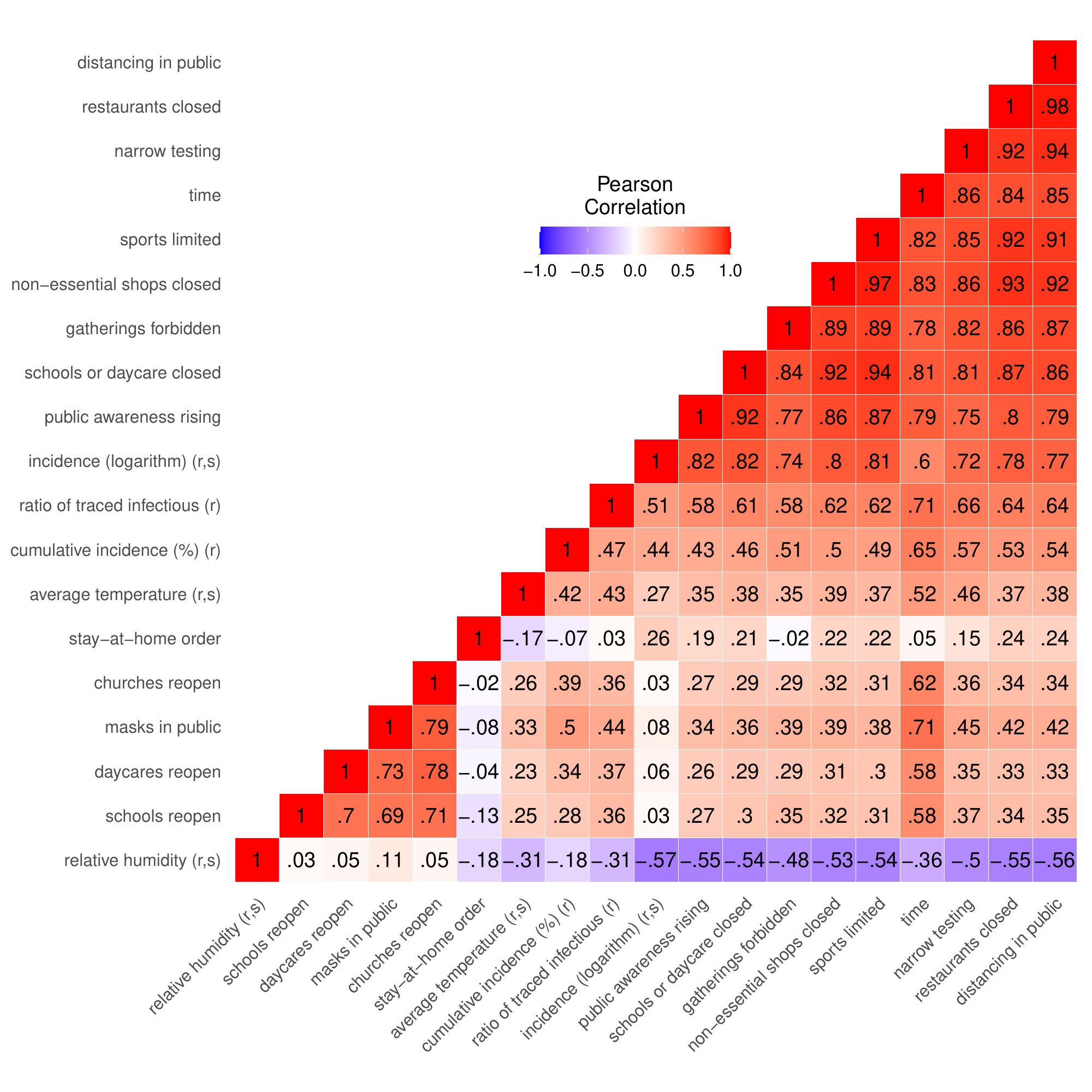}
	\caption{Empirical correlation matrix for most important covariates. As additional variable time was added.}
	\label{fig:cor_main}
\end{figure}

\begin{figure}
	\centering
	\includegraphics[width=\linewidth]{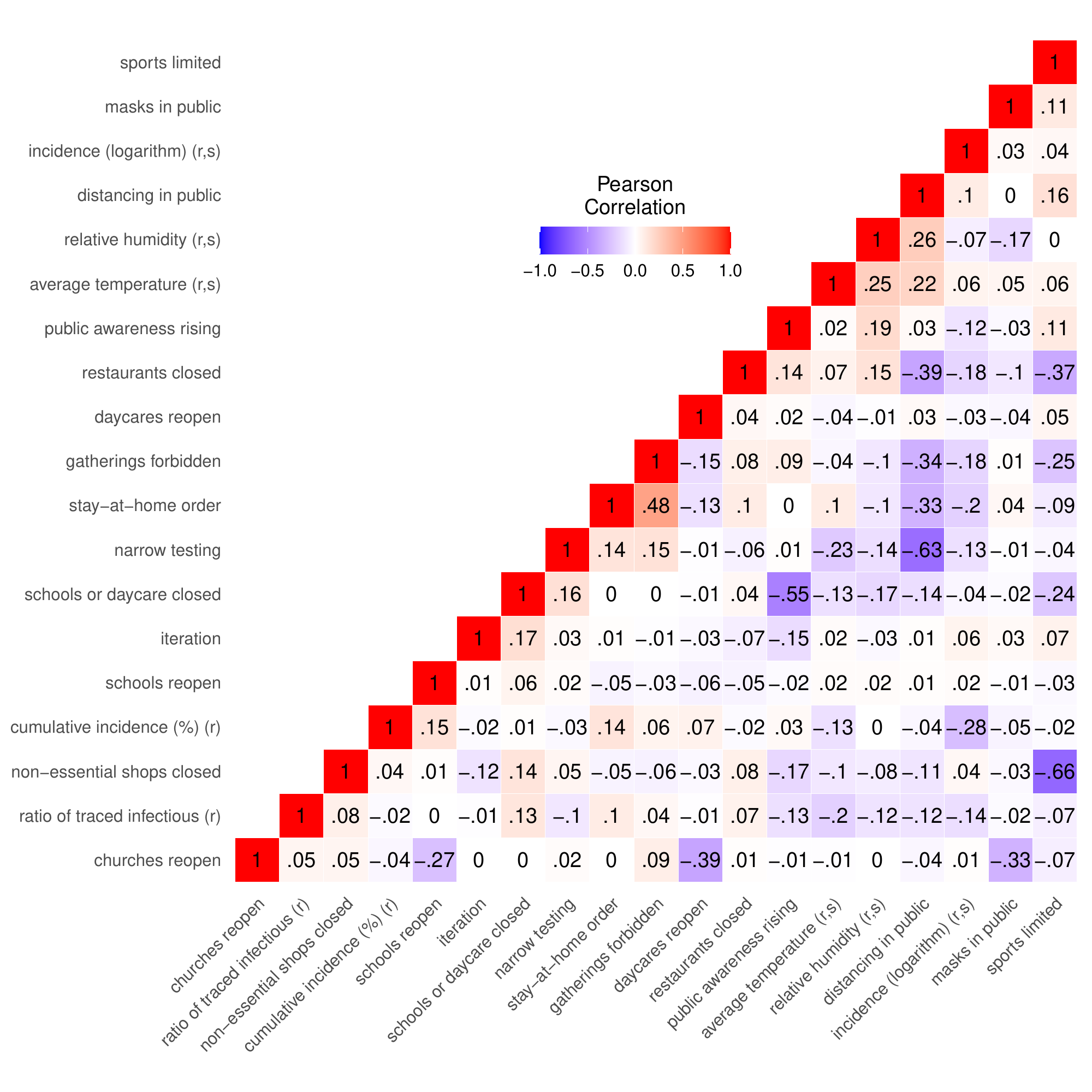}
	\caption{Correlation matrix for effect estimates (averaged across age) of most important variables.}
	\label{fig:estimates_corr}
\end{figure}

As a summary for the reader, Figure \ref{fig:cor_main} illustrates the correlation matrix of the main covariates and Figure \ref{fig:estimates_corr} the correlation matrix of the effect estimates of the main covariates. 

\subsection{Interaction effects}

It is assumed that each covariate increases/decreases a share of the secondary transmissions at a particular day. This ignores interaction effects, which are likely to be of high importance for many interactions. Effect estimates for policy interventions should be understood as estimating the association between the intervention and transmission under the circumstances it was implemented. External validity to other circumstances is subject to discussion.

\subsection{Constant characteristics}

Throughout, the model presented here assumes that properties of the infection are time-independent, in an effort to recover time dependent dynamics of the instantenous reproductive number. This is a simplification. There is for example evidence, that the generation time distribution may be shortened by policy interventions \cite{ali2020serial}. Further, most interventions can be argued to reduce the risk of superspreading events, thereby reducing dispersion. Estimates of dispersion and generation should be understood as empirical averages in the considered time period.

\bibliographystyle{plain}
\bibliography{scibib}

\newpage
\section{Additional figures and tables}\label{supp:additional}

\begin{figure}
	\centering
	\includegraphics[width=\linewidth]{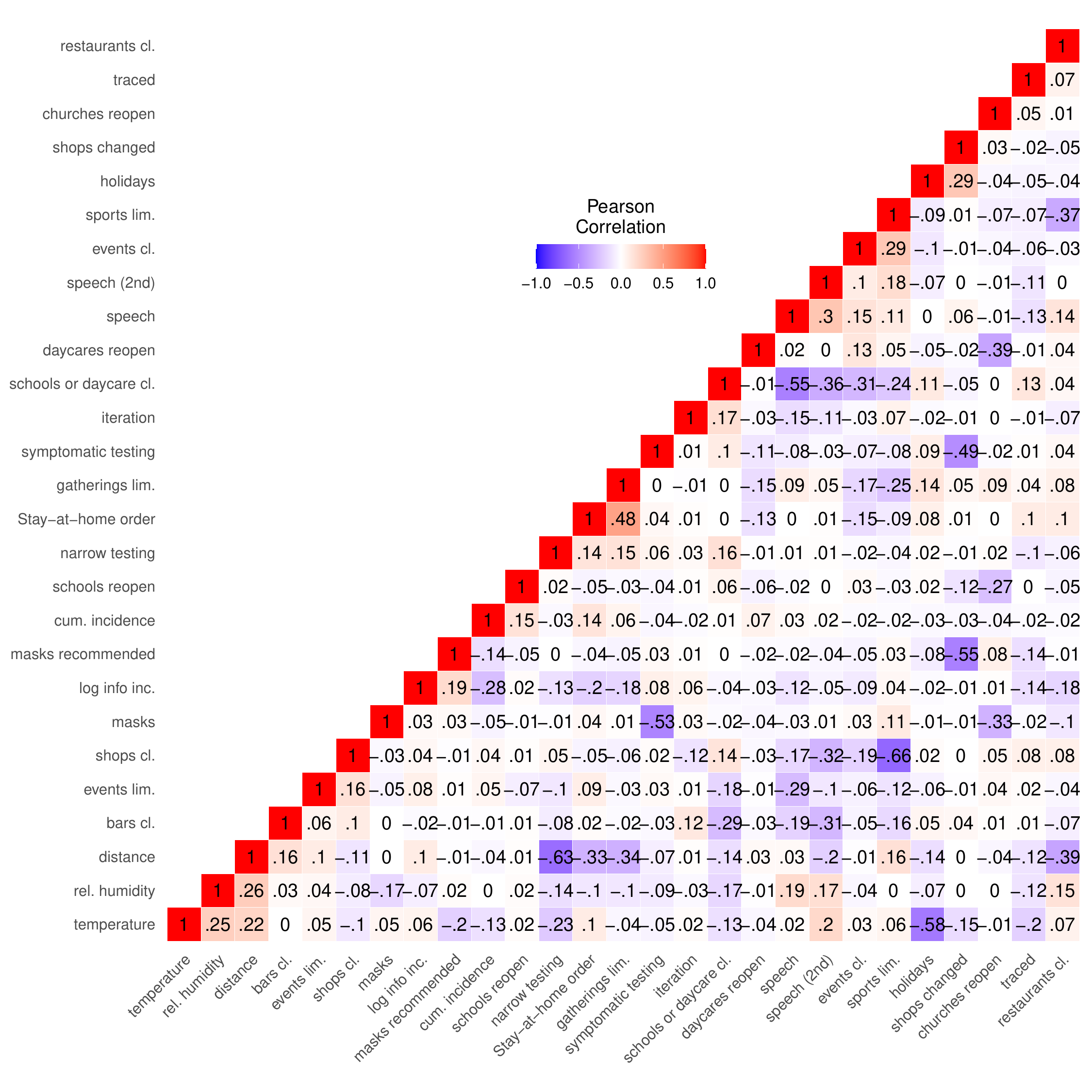}
	\caption{Correlation matrix for effect estimates (averaged across age) for all variables excluding fixed effects.}
	\label{fig:estimates_corr_all}
\end{figure}

\newpage

\small
\begin{longtable}{llrrrrr}
	\caption{List of counties. Only lists cases with symptoms. The symptomatic case fatality rate is denoted by scfr.}\label{tab:counties}\\
	\hline
	name & state & cases & deaths & population (in 1k) & incidence (in 100k) & scfr (\textperthousand) \\ 
	\hline
	Berlin & Ber & 5389 & 173 & 3292 & 164 & 32 \\ 
	LK Altötting & Bay & 468 & 42 & 106 & 441 & 90 \\ 
	LK Amberg-Sulzbach & Bay & 299 & 22 & 104 & 288 & 74 \\ 
	LK Ammerland & Nie & 143 & 3 & 118 & 121 & 21 \\ 
	LK Barnim & Bra & 346 & 27 & 172 & 201 & 78 \\ 
	LK Bayreuth & Bay & 354 & 25 & 105 & 336 & 71 \\ 
	LK Borken & Nor & 704 & 34 & 364 & 194 & 48 \\ 
	LK Breisgau-Hochschwarzwald & Bad & 975 & 66 & 247 & 394 & 68 \\ 
	LK Calw & Bad & 504 & 18 & 151 & 334 & 36 \\ 
	LK Coburg & Bay & 174 & 15 & 88 & 198 & 86 \\ 
	LK Coesfeld & Nor & 584 & 22 & 215 & 271 & 38 \\ 
	LK Cuxhaven & Nie & 131 & 14 & 199 & 66 & 107 \\ 
	LK Dachau & Bay & 707 & 27 & 139 & 510 & 38 \\ 
	LK Dahme-Spreewald & Bra & 164 & 7 & 160 & 102 & 43 \\ 
	LK Diepholz & Nie & 241 & 20 & 211 & 114 & 83 \\ 
	LK Ebersberg & Bay & 360 & 5 & 128 & 282 & 14 \\ 
	LK Eichsfeld & Thü & 123 & 10 & 102 & 120 & 81 \\ 
	LK Euskirchen & Nor & 296 & 19 & 188 & 157 & 64 \\ 
	LK Freising & Bay & 845 & 39 & 163 & 519 & 46 \\ 
	LK Freudenstadt & Bad & 489 & 33 & 116 & 422 & 67 \\ 
	LK Fulda & Hes & 232 & 7 & 216 & 107 & 30 \\ 
	LK Fürstenfeldbruck & Bay & 582 & 22 & 202 & 289 & 38 \\ 
	LK Goslar & Nie & 213 & 24 & 140 & 152 & 113 \\ 
	LK Gotha & Thü & 209 & 28 & 136 & 153 & 134 \\ 
	LK Grafschaft Bentheim & Nie & 152 & 9 & 133 & 114 & 59 \\ 
	LK Greiz & Thü & 502 & 46 & 105 & 477 & 92 \\ 
	LK Groß-Gerau & Hes & 354 & 10 & 251 & 141 & 28 \\ 
	LK Gütersloh & Nor & 583 & 19 & 350 & 166 & 33 \\ 
	LK Havelland & Bra & 154 & 6 & 152 & 101 & 39 \\ 
	LK Heidenheim & Bad & 362 & 35 & 128 & 283 & 97 \\ 
	LK Helmstedt & Nie & 70 & 0 & 91 & 77 & 0 \\ 
	LK Hildesheim & Nie & 291 & 6 & 278 & 105 & 21 \\ 
	LK Hof & Bay & 281 & 13 & 99 & 283 & 46 \\ 
	LK Hohenlohekreis & Bad & 711 & 42 & 107 & 663 & 59 \\ 
	LK Höxter & Nor & 266 & 14 & 146 & 183 & 53 \\ 
	LK Ilm-Kreis & Thü & 96 & 6 & 110 & 87 & 62 \\ 
	LK Kelheim & Bay & 289 & 24 & 113 & 256 & 83 \\ 
	LK Landshut & Bay & 456 & 17 & 147 & 310 & 37 \\ 
	LK Lichtenfels & Bay & 164 & 7 & 67 & 244 & 43 \\ 
	LK Main-Kinzig-Kreis & Hes & 485 & 29 & 401 & 121 & 60 \\ 
	LK Märkisch-Oderland & Bra & 132 & 3 & 187 & 70 & 23 \\ 
	LK Miesbach & Bay & 401 & 8 & 94 & 428 & 20 \\ 
	LK Mühldorf a.Inn & Bay & 351 & 20 & 106 & 330 & 57 \\ 
	LK Nürnberger Land & Bay & 437 & 38 & 164 & 267 & 87 \\ 
	LK Oberhavel & Bra & 224 & 8 & 201 & 112 & 36 \\ 
	LK Odenwaldkreis & Hes & 282 & 45 & 97 & 292 & 160 \\ 
	LK Offenbach & Hes & 421 & 28 & 332 & 127 & 67 \\ 
	LK Oldenburg & Nie & 127 & 4 & 125 & 102 & 31 \\ 
	LK Olpe & Nor & 542 & 48 & 136 & 397 & 89 \\ 
	LK Osnabrück & Nie & 950 & 42 & 350 & 271 & 44 \\ 
	LK Ostalbkreis & Bad & 600 & 28 & 307 & 196 & 47 \\ 
	LK Ostallgäu & Bay & 404 & 29 & 134 & 302 & 72 \\ 
	LK Potsdam-Mittelmark & Bra & 346 & 31 & 203 & 171 & 90 \\ 
	LK Rems-Murr-Kreis & Bad & 822 & 68 & 407 & 202 & 83 \\ 
	LK Reutlingen & Bad & 1129 & 68 & 273 & 413 & 60 \\ 
	LK Rhein-Erft-Kreis & Nor & 763 & 51 & 452 & 169 & 67 \\ 
	LK Rhein-Sieg-Kreis & Nor & 837 & 30 & 578 & 145 & 36 \\ 
	LK Rosenheim & Bay & 1592 & 167 & 244 & 652 & 105 \\ 
	LK Rottal-Inn & Bay & 516 & 24 & 117 & 442 & 47 \\ 
	LK Rottweil & Bad & 616 & 25 & 136 & 451 & 41 \\ 
	LK Saale-Orla-Kreis & Thü & 123 & 12 & 86 & 143 & 98 \\ 
	LK Schmalkalden-Meiningen & Thü & 90 & 3 & 128 & 71 & 33 \\ 
	LK Schwäbisch Hall & Bad & 786 & 57 & 186 & 422 & 73 \\ 
	LK Schwalm-Eder-Kreis & Hes & 365 & 26 & 182 & 201 & 71 \\ 
	LK Sigmaringen & Bad & 666 & 30 & 128 & 522 & 45 \\ 
	LK Sonneberg & Thü & 150 & 14 & 59 & 254 & 93 \\ 
	LK Starnberg & Bay & 263 & 8 & 127 & 207 & 30 \\ 
	LK Steinfurt & Nor & 1247 & 84 & 434 & 288 & 67 \\ 
	LK Straubing-Bogen & Bay & 371 & 25 & 96 & 387 & 67 \\ 
	LK Tirschenreuth & Bay & 950 & 118 & 75 & 1269 & 124 \\ 
	LK Traunstein & Bay & 986 & 71 & 169 & 585 & 72 \\ 
	LK Tübingen & Bad & 1103 & 50 & 212 & 520 & 45 \\ 
	LK Tuttlingen & Bad & 402 & 18 & 132 & 304 & 45 \\ 
	LK Verden & Nie & 103 & 1 & 132 & 78 & 10 \\ 
	LK Werra-Meißner-Kreis & Hes & 106 & 8 & 102 & 104 & 75 \\ 
	LK Wolfenbüttel & Nie & 143 & 11 & 121 & 118 & 77 \\ 
	LK Wunsiedel i.Fichtelgebirge & Bay & 591 & 41 & 76 & 777 & 69 \\ 
	LK Zollernalbkreis & Bad & 890 & 70 & 186 & 480 & 79 \\ 
	Region Hannover & Nie & 1296 & 64 & 1102 & 118 & 49 \\ 
	SK Bonn & Nor & 475 & 7 & 306 & 155 & 15 \\ 
	SK Braunschweig & Nie & 284 & 12 & 243 & 117 & 42 \\ 
	SK Delmenhorst & Nie & 32 & 3 & 73 & 44 & 94 \\ 
	SK Duisburg & Nor & 938 & 56 & 488 & 192 & 60 \\ 
	SK Frankfurt am Main & Hes & 1175 & 54 & 668 & 176 & 46 \\ 
	SK Freiburg i.Breisgau & Bad & 839 & 71 & 210 & 400 & 85 \\ 
	SK Gera & Thü & 147 & 12 & 96 & 153 & 82 \\ 
	SK Hamburg & Ham & 3315 & 139 & 1707 & 194 & 42 \\ 
	SK Hamm & Nor & 350 & 25 & 176 & 199 & 71 \\ 
	SK Heilbronn & Bad & 286 & 5 & 116 & 246 & 17 \\ 
	SK Ingolstadt & Bay & 368 & 30 & 125 & 295 & 82 \\ 
	SK Jena & Thü & 133 & 3 & 106 & 126 & 23 \\ 
	SK Kassel & Hes & 270 & 7 & 191 & 142 & 26 \\ 
	SK Köln & Nor & 1886 & 57 & 1006 & 188 & 30 \\ 
	SK Krefeld & Nor & 455 & 8 & 222 & 205 & 18 \\ 
	SK Landshut & Bay & 192 & 7 & 64 & 302 & 36 \\ 
	SK Leverkusen & Nor & 191 & 4 & 159 & 120 & 21 \\ 
	SK Mönchengladbach & Nor & 471 & 29 & 255 & 185 & 62 \\ 
	SK München & Bay & 3386 & 165 & 1348 & 251 & 49 \\ 
	SK Münster & Nor & 510 & 10 & 290 & 176 & 20 \\ 
	SK Osnabrück & Nie & 448 & 11 & 154 & 291 & 25 \\ 
	SK Pforzheim & Bad & 196 & 7 & 114 & 171 & 36 \\ 
	SK Potsdam & Bra & 444 & 39 & 156 & 285 & 88 \\ 
	SK Regensburg & Bay & 430 & 7 & 135 & 318 & 16 \\ 
	SK Rosenheim & Bay & 328 & 20 & 59 & 553 & 61 \\ 
	SK Salzgitter & Nie & 123 & 8 & 99 & 124 & 65 \\ 
	SK Schwerin & Mec & 46 & 0 & 91 & 50 & 0 \\ 
	SK Straubing & Bay & 277 & 30 & 44 & 623 & 108 \\ 
	SK Wiesbaden & Hes & 309 & 9 & 269 & 115 & 29 \\ 
	SK Wolfsburg & Nie & 163 & 29 & 120 & 136 & 178 \\ 
	SK Wuppertal & Nor & 576 & 40 & 343 & 168 & 69 \\ 
	SK Würzburg & Bay & 310 & 47 & 124 & 249 & 152 \\ 
	\hline
\end{longtable}

\makeatletter\@input{xx.tex}\makeatother